%% file: main.tex
\documentclass[sigplan,screen]{acmart}
\AtBeginDocument{%
  }

\acmYear{2024}\copyrightyear{2024}
\setcopyright{rightsretained}
\acmConference[SOSP '24]{ACM SIGOPS 30th Symposium on Operating Systems Principles}{November 4--6, 2024}{Austin, TX, USA}
\acmBooktitle{ACM SIGOPS 30th Symposium on Operating Systems Principles (SOSP '24), November 4--6, 2024, Austin, TX, USA}
\acmDOI{10.1145/3694715.3695955}
\acmISBN{979-8-4007-1251-7/24/11}

\ccsdesc[500]{Software and its engineering~Compilers}
\ccsdesc[500]{Computer systems organization~Parallel architectures}
\ccsdesc[500]{Computer systems organization~Neural networks}

\keywords{Deep Learning Compiler, Intelligence Processing Unit, Distributed Shared Memory, ML Accelerator}

\usepackage{epsfig,endnotes}
\usepackage{def}
\usepackage{enumitem}
\usepackage{algpseudocode}
\usepackage{graphicx}
\usepackage{textcomp}
\usepackage{fancyhdr}
\usepackage{hyperref}
\usepackage{subcaption}
\usepackage{multirow}
\usepackage{soul}
\usepackage{tikz}
\usepackage{bm}
\usepackage{def}
\usepackage{cleveref}
\usepackage{afterpage}
\usepackage[ruled, vlined, linesnumbered]{algorithm2e}
\usepackage{listings}
\usepackage{pifont}
\usepackage{balance}
\usepackage{tablefootnote}
\usepackage{makecell}
\usepackage{setspace}
\usepackage{soul}
\usepackage{color}

\crefname{section}{\S}{\S}
\newlength{\oldtextfloatsep}
\def\pname{{T10}}
\newcommand{\oursys}{{{T10}}}
\newcommand{\rtensor}{{\textit{r}Tensor}}

\input{author}

\begin{document}

% \pagenumbering{gobble} % stops page numbering throughout document

\title{Scaling Deep Learning Computation over the Inter-core Connected Intelligence Processor with \pname{}} % HotCRP compatible: -0.8ex
\input{abstract}

% \settopmatter{printfolios=true}
% \input{abstract}

\maketitle
% \pagestyle{plain}

%%%%%%%%%%%%%%%%%%%%%%%%%%%%%%%%%%%%%%%%%%%%%%%%%%%%%%%%%%%%%%%%

% 12 pages
\input{intro_new}

\input{background_new}
\input{design_new}

\input{implement}
\input{evaluation}
\input{discussion}
\input{related}

\input{conclusion}
\input{acknowledgment}

% \newpage

% \input{appendix}

%%%%%%%%%%%%%%%%%%%%%%%%%%%%%%%%%%%%%%%%%%%%%%%%%%%%%%%%%%%%%%%%

%% The next two lines define the bibliography style to be used, and
%% the bibliography file.
\newpage
\balance
\bibliographystyle{ACM-Reference-Format}
\bibliography{bib}
\end{document}

%% file: author.tex
\author{Yiqi Liu}
\email{yiqiliu2@illinois.edu}
\affiliation{%
  \institution{UIUC}
  \city{}
  \state{}
  \country{}
}

\author{Yuqi Xue}
\email{yuqixue2@illinois.edu}
\affiliation{%
  \institution{UIUC}
  \city{}
  \state{}
  \country{}
}

\author{Yu Cheng}
\email{yu.cheng@pku.edu.cn}
\affiliation{%
  \institution{Microsoft Research}
  \city{}
  \state{}
  \country{}
}

\author{Lingxiao Ma}
\email{lingxiao.ma@microsoft.com}
\affiliation{%
  \institution{Microsoft Research}
  \city{}
  \state{}
  \country{}
}

\author{Ziming Miao}
\email{ziming.miao@microsoft.com}
\affiliation{%
  \institution{Microsoft Research}
  \city{}
  \state{}
  \country{}
}

\author{Jilong Xue}
\email{jxue@microsoft.com}
\affiliation{%
  \institution{Microsoft Research}
  \city{}
  \state{}
  \country{}
}

\author{Jian Huang}
\email{jianh@illinois.edu}
\affiliation{%
  \institution{UIUC}
  \city{}
  \state{}
  \country{}
}

%% file: abstract.tex
\begin{abstract}
\label{sec:abstract}
As AI chips incorporate numerous parallelized cores to scale deep learning (DL) computing,
inter-core communication is enabled recently 
by employing high-bandwidth
and low-latency
interconnect links on the chip (e.g., Graphcore IPU). It allows each 
core to directly access the fast scratchpad memory in other cores, which enables new parallel computing paradigms.  
However, without proper support for
the scalable
inter-core connections in current DL compilers, 
it is hard for developers to exploit the benefits of this new architecture. 

We present \pname{}, the first DL compiler to exploit the 
inter-core communication bandwidth and distributed on-chip memory on
AI chips. 
To formulate the computation and communication patterns of tensor operators in this new architecture, \pname{} introduces a distributed tensor abstraction \rtensor{}. 
\pname{} maps a DNN model to
execution plans with a generalized compute-shift pattern,
by partitioning DNN computation into sub-operators and mapping them to cores, so that the cores can exchange data following predictable patterns. 
\pname{} makes globally optimized trade-offs between on-chip memory consumption and inter-core communication overhead,
selects the best execution plan from a vast optimization space, and alleviates unnecessary inter-core communications.  
Our evaluation with a real inter-core connected AI chip, the Graphcore IPU, 
shows up to 3.3$\times$ performance improvement, and scalability support for larger models, 
compared to state-of-the-art DL compilers and vendor libraries. 

\end{abstract}

%% file: intro_new.tex
% \vspace{-0.3em}
\section{Introduction}
\label{sec:intro}
% \vspace{-0.1em}

% 1. DNN model like LLM introduce huge running cost.
% 2. Many AI chips explore more scalable and efficient architecture. Inter-core connection a promising direction to further increase HW spec, e.g., TFLOPS and mem BW. E.g., GraphCore IPU provides comparable TFLOPs with V100, but xx times higher memory BW.
% 3. Despite the advantageous HW spec, we found the hardware utilization is extremely low on such accelerators. E.g., the compute utilization is only xx\%, and the memory can only fit a xx\% model size.
% 4. Such limitation is fundamentally caused by .... Introduce the challenges of mapping DNN to IPU.
% 5. Our approach and design insight.
% 6. This approach can benefit to a general class of AI chips.
% 7. evaluation results.

\begin{figure*}[t]
    \centering
    % \vspace{-1.8ex} % hotCRP compatible, do not change: -1.8ex width=0.92\linewidth
    % \includegraphics[width=1\linewidth]{figures/ipu-compare_arch_with_offchip.drawio.pdf}
    % \vspace{-0.5ex}
    \includegraphics[width=0.95\linewidth]{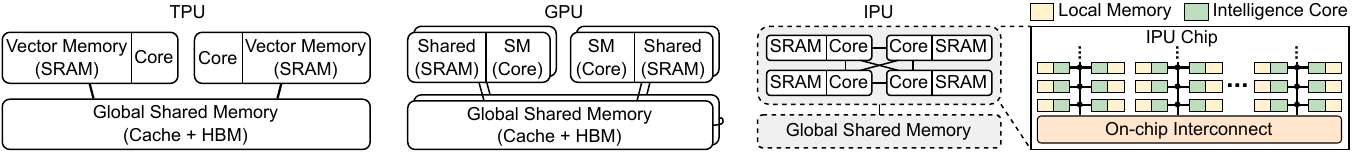} % original
    \vspace{-1.5ex}
    \caption{System architecture of TPU (left), GPU (middle), and IPU (right) chips.}
    \label{fig:compare_arch}
    % \vspace{-0.5ex}
\end{figure*}

% 1. Deep learning workloads are popular today.
% 2. Many hardware accelerators have been developed to support the increasing computation demand of deep learning workloads.
% 3. As the deep learning accelerators continue to evolve, the community is actively exploring new architectural innovations to further improve the performance of deep learning workloads.
% 4. one popular trend is introducing fast inter-core connections that allow data to be exchanged across the compute cores directly without going through the off-chip memory system.

%Over the years, deep learning has made significant progress and revolutionized many disciplines, such as computer vision and natural language processing.
To meet the ever-increasing compute demand of deep learning (DL) workloads, various AI chips or intelligence processors
have been developed~\cite{a100, h100, tpu_supercomputer}.
Typically, an AI chip
%\cite{a100, h100, tpu_supercomputer, ipu2, samba, tenstorrent, wse2}
employs numerous cores 
%such as streaming multi-processors (SMs) on GPUs~\cite{a100} and intelligence processing units (IPUs) on Graphcore~\cite{ipu2}, 
to provide high compute throughput. 
Each core has a small SRAM as its local scratchpad memory. %(e.g., shared memory in each SM in a GPU) 
% which feeds the compute units with data. 
To exploit the parallelism across cores, DL compilers partition the computation into 
multiple pieces.
To synchronize data across cores, all cores share a global memory backed by a high-bandwidth off-chip memory (e.g., HBM). % TODO: rm L2
%To explore the high parallelism, 

%that are computed by different cores in parallel.

%{However, the scalability bottleneck of the global shared memory prevents us from scaling the parallel computing by compacting more cores on a chip.}
Unfortunately, the global memory bandwidth growth gradually lags behind the fast growth of computing performance.
% Instead of relying on the global memory to exchange data among cores, the high-bandwidth and low-latency inter-core communication links have been employed to enable direct data exchange between cores in AI chips. 
Instead of fetching all data from the global memory, inter-core communication links allow cores to directly reuse the data from each other, enabling higher on-chip data reuse.
For example, unlike the TPU~\cite{tpu_v4i} and GPU~\cite{h100} architectures shown in \Cref{fig:compare_arch}, 
the Graphcore IPU~\cite{ipu2} allows each core to access another core's local memory at 5.5GB/s.
The 1,472 cores per chip yield an all-to-all transfer bandwidth of 8TB/s, much higher than the HBM 
bandwidth (1.94TB/s on an A100 GPU).
The aggregated bandwidth can further scale when future technology fits more cores in a chip,
making it a promising approach for breaking the memory bandwidth wall.
Thus, the inter-core links are employed in many emerging 
accelerators, including Graphcore IPU~\cite{ipu2}, SambaNova SN10~\cite{samba}, Cerebras WSE~\cite{wse2}, and Tenstorrent Grayskull~\cite{tenstorrent}.

However, this new architecture
% dramatically
makes executing DL models more complex. 
% also increases the complexity of executing DL models. 
In the traditional \textit{global shared-memory architecture}, programs access all data from a unified memory, so compilers only need to
focus on partitioning computation among cores.
% However, the new architecture requires a coordinated consideration among computation partition, data placement, and inter-core communication. 
In contrast, the on-chip inter-core links enable a \textit{distributed on-chip memory} architecture from programmers' perspective, which requires compilers to coordinate the computation partitioning, data placement, and inter-core communication.
%Due to the lack of such a support in current DL compilers and even vendor runtime libraries (see details in $\S$\ref{sec:limit_existing_approach}), we cannot unleash the full potential of inter-core communication capabilities. 
Simply employing existing compiler techniques causes
% significant 
unnecessary inter-core communication and data duplication in the precious on-chip memory (see \S\ref{sec:limit_existing_approach} and Figure~\ref{fig:save_mem}).
As a result, emerging inter-core connected AI chips fail to compete with the traditional global shared-memory-based AI accelerators due to the inefficient compiler support, although the industry has invested hundreds of millions of dollars in hardware development~\cite{graphcore-investment,sambanova-investment,cerebras-investment}. 

In this paper, we present \oursys{}, a DL compiler for efficiently utilizing
inter-core connected AI chips like Graphcore IPU. 
\oursys{} scales DL computation with distributed on-chip memory by generating 
% a new parallel computing paradigm named ``compute-shift'' 
execution plans with a ``compute-shift'' paradigm.
% By exploiting the deterministic computation patterns of DL workloads, 
\oursys{} carefully plans the tensor placement and transmission among cores
% in an automated manner,
%This enables different DNN models to
% transparently exploit the scalability benefit of the inter-core communication architecture. 
to best utilize both inter-core connection bandwidth and on-chip memory capacity.
First, \oursys{} introduces a distributed tensor abstraction called \textit{RotatingTensor}, or \rtensor{}, that rotates its partitioned tiles across the cores. % \hl{at a configurable pace}.
\oursys{} aligns the rotating paces of different tensors in an operator based on its computation logic and partitioning plan, 
so that data tiles and computation meet at the right timing in every step.
\rtensor{} leverages the compute-shift paradigm to formulate the inter-core communication patterns of DNN workloads, 
which differs from the traditional ``\textit{load-compute-store}'' paradigm~\cite{roller, welder:osdi23} (i.e., load data from global shared memory, compute, and store the results back) used in global shared-memory architectures.

Second, as different tensor partitioning and rotating plans create different trade-offs among the memory footprint, computation time, and communication cost, \oursys{} constructs an accurate cost model to guide the optimization process. 
The compute-shift computing paradigm facilitates the development of an accurate cost model, as it inherently avoids non-deterministic data accesses and allows software to explicitly manage the inter-core data transfers. 
\oursys{} further eliminates the unpredictability in its cost model by exploiting the deterministic computation patterns of DNN workloads.

Third, to flexibly trade-off between memory footprint and communication overhead, \oursys{} builds a spatial-temporal optimization space, where it divides 
an operator into multiple multi-step sub-operators and maps them to individual cores.  
\oursys{} employs a cost-aware operator scheduling process to generate a range of execution plans with varying memory requirements. 
In consideration of the holistic model scheduling, it allows each operator to switch 
between memory- or compute-efficient plans during execution.
\oursys{} then applies a holistic 
reconciliation process to search for an optimized end-to-end execution plan that can fit multiple operators or 
even the entire model into the distributed on-chip memory. 
%that can fit the entire model within a given device memory capacity. 
%After that, with a set of abstracted device interfaces, \pname{} translates the execution plan into a specific execution program, consisting of nested loops over computation and shift-based communication stages across all cores.

We implement \oursys{} in 10K lines of code with Python, C/C++, and assembly, for the execution plan optimization and kernel code generation.
We evaluate \oursys{} with various DNN models, including CNNs and transformers with varying batch sizes. The models are executed on a real Graphcore IPU MK2 chip.
% , as well as its variants by configuring its available cores and memory capacity.
% As the IPU chip is not backed by off-chip memory, \oursys{} will attempt to fit the entire DNN model into the on-chip memory.
\oursys{} outperforms current DL compilers and vendor libraries by up to 3.3$\times$, 
and allows much larger models and batch sizes to fit into the distributed on-chip memory.
%e.g., with up to a \todo{xxx} improvement. 
We also show its benefits in serving large language models. Overall, we make the following contributions in this paper:

\begin{itemize}[topsep=0em,itemsep=0em,parsep=0em,partopsep=0em,leftmargin=*]
    \vspace{1ex}
    % \item We develop \oursys{}, the first deep learning compiler to exploit the massive inter-core communication bandwidth on distributed on-chip memory accelerators.
    \item We develop \oursys{}, the first DL compiler to exploit the inter-core communication bandwidth and best utilize distributed on-chip memory on intelligence processors (\S\ref{sec:coreidea}).
    %\vspace{-2ex}
    % \item We propose the \rtensor{} abstraction, which generalizes the classic distributed matrix multiplication approach (i.e., Cannon's algorithm~\cite{cannon}) for general tensor computation.
    \item We propose a new tensor abstraction to formulate the inter-core transfer patterns of DNN models and represent them with the compute-shift execution plan (\S\ref{sec:design:rtensor} \& \S\ref{sec:design:compute}).
    % systematic compiler approach and optimization space

    %%%%%% These two items:
    
    \item We build an accurate cost model to guide the optimization for computation partitioning and inter-core communications, significantly reducing the compilation time (\S\ref{sec:design:optimization}).
   
    \item We propose a cost-aware operator scheduling process to tradeoff memory footprint and communication overhead, and generate optimized end-to-end execution plan (\S\ref{sec:design:optimization}).

    %%%%%% Or this one item:
    
    % \item \hl{We propose a cost-aware scheduling process to optimize the operator partitioning and inter-core data transfer, tradeoff the memory footprint and communication overhead, and generate an optimized end-to-end execution plan} \S\ref{sec:design:optimization}).

    \item We implement \oursys{} as a standalone compiler and develop generic device interface for enabling the mapping of tensor operators to inter-core connected AI chips (\S\ref{sec:design:hardware} \& \S\ref{sec:implementation}).
    
    \item We evaluate \oursys{} with a real Graphcore IPU MK2 chip and show its efficiency and scalability for DNN workloads (\S\ref{sec:evaluation}).
\end{itemize}

% \begin{figure}
%     \centering
%     % \includegraphics[scale=0.62]{figures/ipu-ipu_arch_no_offchip.drawio.pdf}
%     \includegraphics[scale=0.62]{figures/ipu-ipu_arch.drawio.pdf} % original figure
%     \vspace{-3ex}
%     \caption{System architecture of an intelligence processing unit (IPU) with cores fully connected by on-chip links.}
%     \label{fig:ipu_arch}
%     \vspace{-5ex}
% \end{figure}

%% file: background_new.tex
% \vspace{-0.2em}
\section{Background and Motivation}
\label{sec:background}
% \vspace{-0.1em}
% In this section, we will first introduce the system architecture of the inter-core connected intelligence processor. After that, we will discuss the motivation for our work. 
We introduce the system architecture of inter-core connected intelligence processor and discuss the motivation of \pname{}.

\begin{figure*}
    \centering
    % \vspace{-2.2ex} % hotCRP compatible, do not change: -2.2ex scale=0.65
    \includegraphics[scale=0.7]{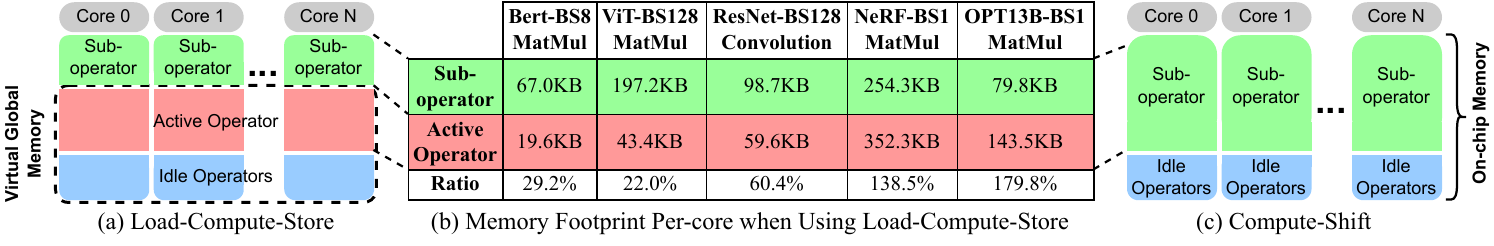}
    \vspace{-5.5ex}
    \caption{A comparison of the conventional load-compute-store (a) vs. our compute-shift (c) style execution.
    (b) shows the per-core memory footprint of representative operators when running DNN models on IPU using VGM. 
    \textbf{Ratio} is the potential increase in sub-operator size by removing VGM.
    The result of OPT13B~\cite{opt} comes from profiling one of its layers on IPU.}
    \label{fig:save_mem}
    \vspace{-1ex}
    % https://app.diagrams.net/#G1gcI2pbnteWPXEvKO05_x6oUnLR3acn6l#%7B%22pageId%22%3A%22EFf7PqpgCWQ3Ouzl8fFr%22%7D
\end{figure*}

%In this section, we first introduce the system architecture of a typical massive-parallel intelligence processor that supports 
%inter-core communications with distributed on-chip memory. After that, we will discuss its programming challenges. 

% \vspace{-0.1em}
% \subsection{Distributed On-chip Memory Architecture}
%\subsection{Intelligence Processing Unit}
\subsection{Inter-core Connected Intelligence Processor}

% \input{figures/table_compare}

%To meet the increasing demand for higher memory throughput in modern DNN computations,
As discussed in $\S$\ref{sec:intro}, many AI chips are adopting the inter-core connected architecture~\cite{ipu2, samba, tenstorrent, wse2}.
In this paper, 
we focus on a representative example, the Graphcore Intelligence Processing Unit (IPU) MK2\mbox{~\cite{ipu2}}, as shown in\mbox{~\Cref{fig:compare_arch}} (right).
An IPU chip has 1,472 cores, and each core executes independent threads in parallel with a private 624KB scratchpad memory, 
which adds up to a total of 896MB on-chip memory.
% Unlike in the global shared memory architecture, the IPU cores are directly interconnected by high-bandwidth and low-latency links, allowing each core to access the scratchpad memory of another core at 5.5GB/s, offering an aggregated inter-core all-to-all bandwidth of $1472\times5.5\mathrm{GB/s}\approx 8\mathrm{TB/s}$\mbox{~\cite{ipu_citadel}}.
Compared to the global shared memory architecture,
a key
%the most notable
distinction is that IPU cores are interconnected by high-bandwidth low-latency links. Each core can access the scratchpad memory of another core at 5.5GB/s, offering an aggregated inter-core all-to-all bandwidth of $1472\times5.5\mathrm{GB/s}\approx 8\mathrm{TB/s}$\mbox{~\cite{ipu_citadel}}.

% The high aggregated memory bandwidth and large on-chip memory capacity imply new optimization opportunities for deep learning (DL) workloads to enable a more scalable parallel computing paradigm.
%The high aggregated memory bandwidth and large on-chip memory capacity enable more scalable performance for serving highly parallelized deep learning (DL) workloads.
Specifically, cores can individually access data from other cores, without contending for the global shared memory. 
%For example, a dataflow programming model where one core directly reads data from another core can be enabled, as opposed to the traditional approach where each core works on the distinct data chunks and synchronizes the results via a global shared memory.
Thus, the inter-core connection effectively aggregates the local memories of all cores into a large distributed on-chip memory with 8TB/s total bandwidth, which is much higher than a global shared memory backed by off-chip HBM (e.g., 1.94TB/s on an A100 GPU).
This can alleviate the well-known memory bandwidth bottleneck in DL applications.

% \begin{figure}
%     \centering
%     \includegraphics[width=0.76\linewidth]{figures/motiv_mem_duplicate.pdf}
%     \vspace{-3ex}
%     \caption{Data duplication ratio of various operators on IPU.}
%     \label{fig:ipu_mem_duplication}
%     \vspace{-2ex}
% \end{figure}

% \vspace{-1em}
\subsection{Inefficiency of Existing Approaches}\label{sec:limit_existing_approach}
% \vspace{-0.1em}

%While enabling impressive memory performance, the distributed shared memory also presents new challenges.
%Unlike the traditional case where all tensor data is accessed from the unified global memory space, the distributed memory requires compilers to explicitly partition each tensor to individual cores and orchestrate all inter-core data transfers.

% Given the challenges of optimizing computation and communication for distributed on-chip memory, 

%Amid the above complexities, 
To support inter-core connected AI chips, 
existing compilers~\cite{ansor,roller} 
and libraries~\cite{popart} mimic a shared memory for all cores by reserving a portion of local memory in each core and 
abstracting them as a ``virtual global memory'' (VGM), as shown in \Cref{fig:save_mem} (a).
% By default, all tensors, including persistent 
% weights (e.g., from idle operators that are not currently running) and temporary activations (e.g., from the active operator that is currently performing computation), are placed in the virtual global memory.  
\hlB{By default, to store an entire DL model on chip, all tensors used by the operators of the model, including persistent weights and temporary activations, are placed in the VGM.}
% During execution, similar to a shared-memory-based accelerator,
During execution, the active operator (i.e., currently running operator) is partitioned into sub-operators, each running on one core.
Tensors of the idle operators (i.e.,
% \hl{pre-loaded}
operators
stored on-chip but
not currently running) are unused in the VGM.
% For the active operator, each core retrieves data from the VGM to its local memory, performs computation locally,
To execute a sub-operator, each core retrieves data from VGM to its local ``sub-operator'' memory (shaded in green in \Cref{fig:save_mem}), performs computation locally, 
and stores the result back to VGM.
We define this as a ``\textit{load-compute-store}'' paradigm. 

%\footnotetext{We run one of the 40 identical layers of the OPT13B model on an IPU chip, as the current generation IPU only has 896MB on-chip memory. Entire model may be executed by connecting multiple chips as a pipeline (see \S\ref{sec:eval_llm}).}

% \bgroup
% \setlength\tabcolsep{2pt}
% \begin{table}
% \centering
% \caption{Per-core memory footprint breakdown of representative operators when running DNN models on IPU using VGM (see \Cref{fig:save_mem} (a)). ``Ratio'' is the potential increase in sub-operator size if we remove the VGM abstraction.}
% \vspace{-2ex}
% \footnotesize
% \begin{tabular}{|l|l|l|l|l|}
% \hline   & \textbf{Active Operator}   & \textbf{Sub-operator} & \textbf{Ratio}\\\hline 
% \textbf{Bert-BS8 MatMul}    & 19.6KB  & 67.0KB                & 29.2\%        \\
% \textbf{Vit-BS128 MatMul}   & 43.4KB  & 197.2KB               & 22.0\%        \\
% \textbf{Resnet-BS128 Conv}  & 59.6KB  & 98.7KB                & 60.4\%        \\
% \textbf{NeRF-BS1 MatMul}    & 352.3KB & 254.3KB               & 138.5\%       \\
% \textbf{OPT13B-BS1 MatMul}     
%                             & 143.5KB & 79.8KB                & 179.8\%       \\\hline
% \end{tabular}
% \label{tab:vgm}
% \vspace{-3ex}
% \end{table}
% \egroup
\begin{figure*}[t]
    \centering
    % \vspace{-1ex} % hotCRP compatible, do not change: -2ex width=0.8\linewidth
    \includegraphics[width=0.93\linewidth]{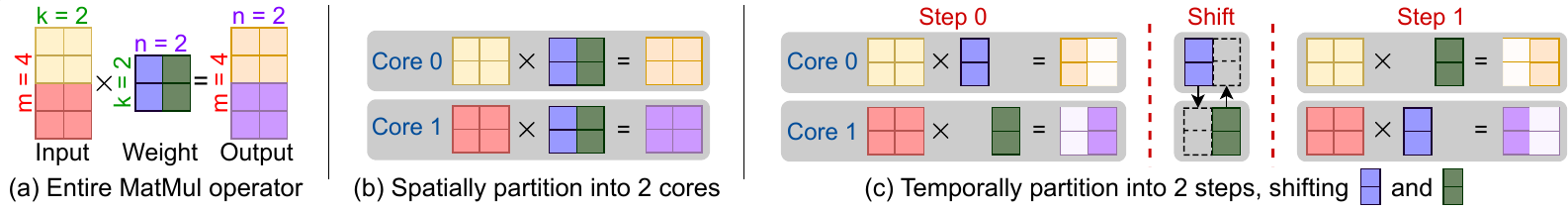}
    \vspace{-2.5ex}
    %\caption{Fit operator into cores with limited capacity using shift, assuming each core can hold 11 square elements.}
    \caption{An example that maps a MatMul operator to two cores with the compute-shift style execution. Both (b) and (c) are valid compute-shift execution plans, but with different tradeoffs between memory footprint and communication overhead.}
    \label{fig:motivation_example}
    \vspace{-1ex}
\end{figure*}

% \vspace{0.1em}
\noindent
\textbf{Inefficient inter-core communications.}
VGM introduces significant inefficiencies in inter-core communications.

First, accessing tensor data from VGM causes \textit{imbalanced memory accesses} across cores, where some cores issue or serve more data accesses than others. 
% Specifically, when the tensor placement is not aligned with the placement of computation tasks, 
As each tensor partition is stored in one core but often used by multiple cores for computation, some cores can obtain needed partitions from their local memory, while other cores need to remotely fetch required partitions from their peers. Then, the execution is bottlenecked by cores that access remotely.
Also, sub-optimal tensor placements may cause bandwidth contentions. For instance, when multiple cores access different data from the same core, these cores will contend for the limited 5.5GB/s bandwidth of a single core, stalling the entire execution.
% imbalanced access may also occur when more data are accessed from some cores. For example, a matrix multiplication (MatMul) operator requires a row of one input tensor and a column of another for each step. Thus, the core that stores the intersection of the row and the column must serve more data accesses, which bottlenecks the entire chip.

%partitioning tensors in VGM can generate many small fragments, 
% \hlcyan{Second, when a tensor is split across different cores via VGM, it generates many small fragments among these cores. Therefore, 
% a core needs to synchronize and communicate with many cores to fetch each piece of the tensor to retrieve a complete one,} causing in \textit{redundant inter-core communications}. %hurting the overall performance. 

Second, to store a tensor using the VGM, as shown by the red ``Active Operator'' boxes in \mbox{\Cref{fig:save_mem} (a)}, we split it into small pieces across multiple cores. To retrieve a complete tensor, a core must fetch each piece from a different core, requiring it to communicate with multiple cores. This leads to
% many inter-core synchronizations and 
\textit{redundant inter-core communications}.

To quantify the inter-core communication overhead, we break down the data transfer and compute time for a modern compiler that uses VGM, such as Roller~\mbox{\cite{roller}}. 
%and optimizes the tile size for operators.
With VGM, the inter-core data accesses account for 50\%--74\% of the end-to-end execution time (see \Cref{fig:compute_shift_breakdown}).
We will show how \pname{} reduces the overhead to 8\%--43\% by eliminating VGM and orchestrating the inter-core communications in $\S$\ref{sec:coreidea} and $\S$\ref{sec:design}.
% \todo{move to eval}

% \vspace{0.1em}
\noindent
\textbf{Inefficient use of on-chip memory.}
The VGM uses the on-chip memory capacity inefficiently.
% In the load-compute-store model, the virtual global memory serves as an inclusive cache, where all data that exists in the local memory must also exist in the global memory.
As shown in Figure~\ref{fig:save_mem} (a), each sub-operator of the currently active operator loads required data from the VGM to 
its local memory, which duplicates the data in both memory spaces.
To host the duplicated data, VGM reserves memory space on each core, as shown by the active operator region in Figure~\ref{fig:save_mem} (a). 
This leaves less free on-chip memory, restricts each core to accommodate a smaller sub-operator, and results in \textit{low compute intensity}. 
With less data reuse inside a core, higher data transfer volume is required for performing the same computation.
% Thus, by using more on-chip memory to store the duplicated data, the virtual global memory increases the inter-core communication volume and further intensifies the overall performance degradation.

%Specifically, the size of duplicated data by virtual global memory is often measured as the total input tensor size of the currently active operator.
We quantify the storage overhead of VGM for representative operators in \mbox{\Cref{fig:save_mem}} (b). 
%Comparing the active operator and sub-operator regions, 
By removing the VGM (i.e., merging the active operator region into the sub-operator region) in \Cref{fig:save_mem} (c), we can increase sub-operator size by 22\%--180\%. \pname{} leverages this to improve memory efficiency.

% For example, a matrix multiplication operator from transformer models like OPT-13B may have a 205MB input tensor~\cite{opt}, while an intelligence processor 
% like IPU has 896MB on-chip memory (see Table~\ref{tab:compare_arch}).
% %and one from other evaluated models may exceed 250MB in input size~\cite{nerf}.
% The data duplication caused by VGM consumes a large portion of on-chip memory. 
% %will impact the performance of intelligence processors like the IPU, which has 896MB of on-chip memory in total.
% As idle operators waiting to be executed also occupy the on-chip memory, the available memory further shrinks. 
% Thus, best utilizing on-chip memory is also an objective of our work.
% %so freeing up the active operator space becomes more urgent.
% % Meanwhile, to continuously exploit the fast on-chip connection and memory, it is common to also place the operators that will be executed later onto the chip in advance (the Idle Operators region in Figure~\ref{fig:save_mem}(a)), which further shrinks the available memory space. 
% % In this case, freeing up the space occupied by duplicated data brings even more impact.
% % Taking IPU as an example, since there is no off-chip HBM on it, the entire DNN model must be placed on-chip to avoid using the extremely slow host memory. By placing a 680

% \vspace{-0.1em}
\section{Core Idea of \pname{}}
\label{sec:coreidea}
% \vspace{-0.1em}
% regular data access observation->rotating tensors
% duplicated tensor can be further partitioned -> 1) no memory waste -> 2) large op tile 
% equal partition -> balanced communication; 
% align computation with communication tile  -> one data transfer at a time

%The inefficiencies of VGM prevent the high aggregated bandwidth and capacity on an IPU chip from unleashing their full potential, disabling a promising opportunity for breaking the memory bandwidth wall.
%To eliminate the VGM, we carefully partition the operator and its tensors to the individual cores and their local memories.
%we organize the computation and inter-core data transfer as \textcolor{red}{multiple nested circular pipelines}.
% By organizing the computation in a dataflow manner, every core computes one data tile at a time by continuously shifting the data tiles across cores.

\begin{figure}[t]
    \centering
    % \vspace{-3ex} % hotCRP compatible, do not change:
    \includegraphics[scale=0.575]{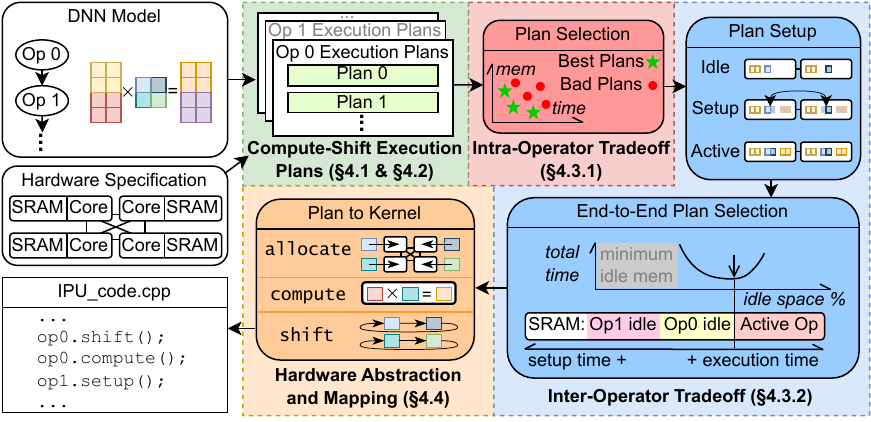}
    \vspace{-5ex}
    \caption{System overview of \pname{}.}
    \label{fig:overview}
    \vspace{-2ex}
\end{figure}

% To share tensors among multiple cores without incurring excessive memory footprint or redundant inter-core communications, 
To eliminate the excessive memory footprint and redundant inter-core communications of VGM, 
we map the DNN computation to a \textit{compute-shift} pattern.
% we adopt a \textit{compute-shift} execution pattern to map and optimize the DNN computation to an inter-core connected AI chip.
In each step, each core independently computes a sub-task with data received from its upstream neighbors and shifts the data to its downstream.
% The feasibility of applying this approach to general DNNs is based on our observation on their computation patterns: 
The feasibility of this approach for general DNNs comes from this observation:
% most DNN operators can be divided into regular computation tasks, which load consecutive data tiles from input tensors and produce consecutive data tiles of the output tensor. 
most DNN operators can be divided into regular computation tasks, which load and produce consecutive data tiles of the input and output tensors, respectively. 

We show an example that maps a matrix multiplication (MatMul) operator to two cores in \Cref{fig:motivation_example} (a).
We first partition the operator along dimension $m$ onto two cores in \Cref{fig:motivation_example} (b). 
By default, both cores hold a copy of the weight tensor, which incurs memory capacity overhead. 
To reduce memory footprint, in \Cref{fig:motivation_example} (c), we further split the weight tensor along dimension $n$ into two parts and place each part on one of the cores.
Then, the computation must be conducted in two steps, as each core holds half of the weight tensor and performs half of its computation per step.
Between the computation steps, each core circularly shifts its partition to the next core, forming a shift ring of two cores.

The compute-shift pattern avoids the inefficiencies of VGM.
% by trading-off between memory footprint and inter-core communication.
First, each part of the weight tensor is stored in at least one core at any time, which eliminates the need for a global memory to store shared data.
This improves the memory capacity utilization and allows larger sub-operator sizes.
% This enables improved memory capacity utilization and larger sub-operator size.
Second, by
% evenly partitioning
circularly shifting 
tensors across cores, the communication volume is evenly distributed across the inter-core connections.
Third, as we accurately align the computation with data tile (e.g., \Cref{fig:motivation_example} (c) shifts a 2$\times$1 weight tile and then computes a 2$\times$1 output tile), each core only needs to communicate with one other core for each tensor at each step, avoiding redundant communications to many cores.

% \vspace{0.5em}
% \noindent\textbf{The tradeoff behind compute-shift.}
To find the best execution plan with the compute-shift pattern, we must exploit \textul{\textit{the tradeoff between memory footprint and communication overhead}}.
For example, both \mbox{\Cref{fig:motivation_example}} (b) and (c) show valid execution plans. Plan (b) finishes the entire computation in one step without inter-core communication, but has a higher memory footprint. Plan (c) has less memory footprint but incurs more communication overhead.
% Thus, we can derive a \textit{tradeoff between memory footprint and communication overhead} for each operator.

% consider the tradeoff on 1000 cores ia complex

% \hl{Also, while various DNN operators have different demands on inter-core communication bandwidth and on-chip memory capacity, these hardware resources are limited.
% % Thus, we must tradeoff carefully on the operators to optimize their overall performance within the resource constraints.

% To tradeoff on various operators and optimize for better performence, we will overcome the following challenges.
In reality, as we consider multi-dimensional DNN operators and thousands of cores on an IPU chip, deriving the best tradeoff can be difficult. 
An efficient compute-shift execution plan for
% a large operator on thousands of cores 
them
may contain numerous nested shift rings along multiple tensor dimensions, composing a massive tradeoff space to search through.
%Moreover, different operators have different tradeoffs for performance.
%(i.e., some prefer more memory footprint while some prefer more communication volume). 
Given limited inter-core connection bandwidth and on-chip memory capacity, we must also holistically tradeoff among multiple operators on the chip to derive an optimized end-to-end execution plan. 

%% file: design_new.tex
\section{System Design of \oursys{}}
\label{sec:design}
% \vspace{-0.5em}

\begin{figure}[t]
    \centering
   % \vspace{-0.5ex}
    \includegraphics[scale=0.55]{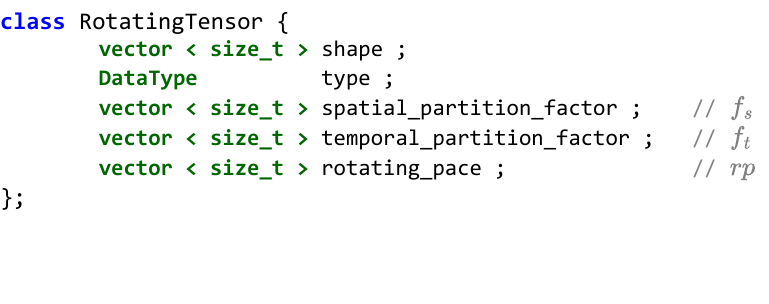}
    \vspace{-8.5ex}
    \caption{\rtensor{} abstraction in \oursys{}.}
    \label{fig:rtensor}
    \vspace{-2ex}
\end{figure}

%To generalize the above idea for various DNN models and accelerators, 
We now introduce \oursys{}, a compiler designed to optimize end-to-end DNN model execution on an inter-core connected intelligence processor.
We present the overview of \pname{} in \Cref{fig:overview}: 
\textbf{(1)} \pname{} introduces the RotatingTensor (\rtensor{}) abstraction to represent the partitioning and communication patterns of tensor operators on distributed on-chip memory ($\S$\ref{sec:design:rtensor}).
\textbf{(2)} It uses \rtensor{} to map a DNN model to compute-shift execution plans.
\hlE{The rich expressiveness of rTensor enables the tradeoff between memory usage and communication overhead} ($\S$\ref{sec:design:compute}).
\textbf{(3)} By configuring \rtensor{}s in different ways, \pname{} defines a comprehensive optimization space for a DNN model. 
It adopts a two-staged optimization strategy to
handle the tradeoff between inter-core communication and memory footprint, and
optimize for end-to-end DNN model execution.
For each operator, \pname{} finds the Pareto-optimal execution plans that represent the best trade-off between the execution time and memory footprint ($\S$\ref{subsubsec:intra-operator}).
Then, \pname{} employs a holistic inter-operator memory reconciliation policy to determine the best 
end-to-end plan for the DNN model ($\S$\ref{subsubsec:inter-operator}).
\textbf{(4)} The plan is compiled onto the processor using three abstracted device interfaces ($\S$\ref{sec:design:hardware}).

\subsection{\rtensor{}: A New Tensor Abstraction}
\label{sec:design:rtensor}

% \todo{intercore connect for general CPU is an old thing, dataflow arch is intrinsically hard to optimize (2.2), not new compute model, highlight constraints and tradeoff, use 4.3}

To map a tensor onto the distributed on-chip memory and stream the partitioned sub-tensors across multiple groups of cores, \oursys{} introduces a distributed tensor abstraction called RotatingTensor (\rtensor{}), as shown in \Cref{fig:rtensor}.
In addition to defining the tensor shape and data type, \rtensor{} also describes how each tensor is partitioned, mapped, and shifted on the interconnected cores (summarized in \Cref{tab:terms}).

\begin{table}[t]
    \centering
   % \vspace{-0.3ex}
    \caption{Terminology used in \pname{}.}
    \vspace{-2.5ex}
    %\footnotesize
    \scriptsize
    \begin{tabular}{|c|c|c|}
    \hline
        \textbf{Symbol} & \textbf{Name} & \textbf{Description} \\\hline
        $f_s^X$ & \makecell{Spatial \\ Partition Factor} & \makecell{Spatially partitions a tensor $X$ \\ into sub-tensors.} \\\hline
        $f_t^X$ & \makecell{Temporal \\ Partition Factor} & \makecell{Temporally partitions a sub-tensor of $X$ \\ into sub-tensor partitions.} \\\hline
        $\rp$ & Rotating Pace & \makecell{Specifies how sub-tensor partitions \\ are shifted among cores.} \\\hline
        % Temporally partition a sub-operator \\ into sub-tasks (steps).
        $F_{op}$ & \makecell{Operator \\ Partition Factor} & \makecell{Spatially partitions an entire operator \\ into sub-operators.} \\\hline
    \end{tabular}
    \label{tab:terms}
    \vspace{-2.5ex}
\end{table}

\begin{figure*}[t]
    \centering
    % \vspace{-3.6ex} % can be very high ???
    \includegraphics[width=1\linewidth]{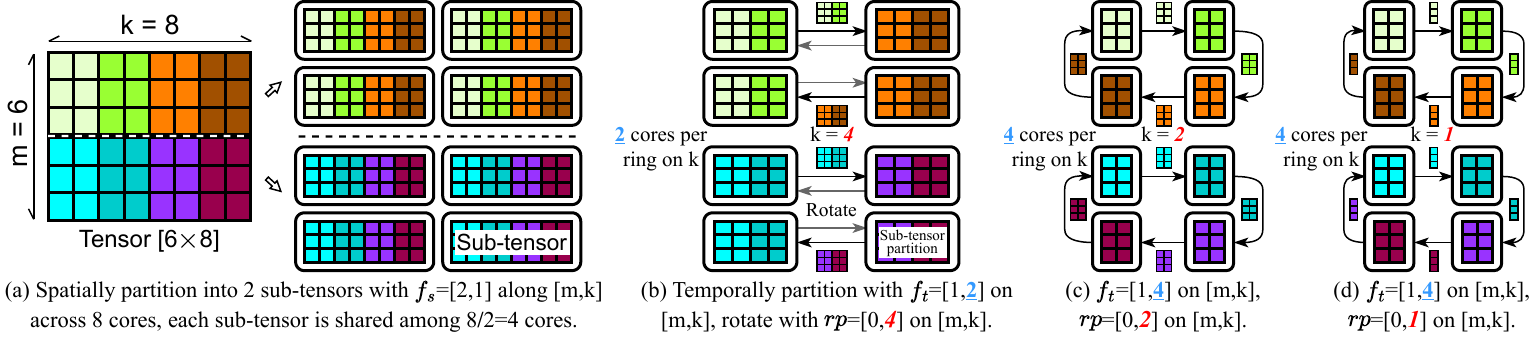}
    \vspace{-6ex}
    \caption{Illustration of \rtensor{} partitioning and rotating.}
    \label{fig:shift}
    \vspace{-2ex}
\end{figure*}

First, \pname{} partitions the computation of an operator onto multiple cores.
Based on the data dependency, the computation partitioning will imply how each of its input/output tensor is partitioned.
This gives a \textit{spatial partition factor} ($\bm{f_s}$), which splits a tensor into \textbf{\textit{sub-tensors}}.
Second, each sub-tensor may be required by multiple cores.
To share a sub-tensor among them, we specify how the sub-tensor is further partitioned among the cores using a \textit{temporal partition factor} ($\bm{f_t}$).
Third, we specify how the partitions of a sub-tensor are circularly shifted among the cores using the \textit{rotating pace} ($\bmrp$).
Altogether, a set of \rtensor{}s of an operator defines a compute-shift execution plan. The numerous possible \rtensor{} configurations of an operator generate a combinatorial optimization space of execution plans.

Specifically, $f_s$, $f_t$, and $\rp$ are vectors with a length equal to the number of dimensions of a tensor, indicating how the tensor is partitioned along each dimension.
For example, in \mbox{\Cref{fig:shift} (a)}, a tensor $T$ of shape $[6, 8]$ is partitioned onto 8 cores by a spatial factor $f_s=[2, 1]$, forming 2 sub-tensors of shape $[3, 8]$.
Thus, to share each sub-tensor among 4 cores without incurring high memory footprint, a temporal factor $f_t=[1,2]$ further partitions each sub-tensor into $2$ partitions with shape $[3, 4]$, as shown in \mbox{\Cref{fig:shift} (b)}.
It forms $\frac{4}{2}=2$ rotation rings with $2$ cores in each, where cores share the sub-tensor by circularly shifting its partitions.
In comparison, \Cref{fig:shift} (c) shows how another $f_t=[1,4]$ splits the same sub-tensor to $4$ partitions, on $\frac{4}{4}$=1 rotation ring with $4$ cores.

Finally, the rotating pace $\rp$ controls how fast an \rtensor{} rotates, so we can align the data shifting with computation (see \S\ref{sec:design:compute}).
Practically, $\rp$ specifies the number of data elements shifted in each step along each tensor dimension. For example, $\rp=[0, 2]$ in \Cref{fig:shift} (c) means that for each step, the sub-tensor shifts 2 elements along the second dimension (i.e., a data tile of shape $[3, 2]$), and finishes a full cycle in $\frac{8}{2}$=4 steps.
Notably, different $\rp$s can be applied to the same set of $f_s$ and $f_t$. For instance, in \Cref{fig:shift} (d), $\rp=[0, 1]$ shifts a tile of $[3, 1]$ for each step, requiring $\frac{8}{1}$=8 steps in total.

% \begin{figure} [t]
% \small
% \centering
% \noindent
% \begin{minipage}[t]{0.9\linewidth}
% \begin{lstlisting}
% class RotatingTensor {
%   vector<size_t> shape;
%   DataType type;
%   vector<size_t> spatial_factor;
%   vector<size_t> temporal_factor;
%   vector<size_t> rotating_pace;
% };
% \end{lstlisting}
% \end{minipage}
% \vspace{-3ex}
% \caption{\rtensor{} abstraction in \oursys{}.} 
% \vspace{-2ex}
% \label{fig:rtensor}
% \end{figure}

\subsection{\textit{Compute-Shift} Execution Plan}
\label{sec:design:compute}
% \vspace{-0.3em}

Using the \rtensor{} abstraction, \oursys{} organizes the computation of a general DNN operator into a compute-shift pattern, where the operator's computation and tensors are partitioned to individual cores and their local memories.
The entire computation involves multiple compute-shift steps until each tensor has been shifted across all cores.
Each compute step is defined as a \textbf{\textit{sub-task}}.
In each compute-shift step, each core computes a sub-task
% (i.e., the computation on each core at each step)
and shifts local tensors to its neighbors.
We now discuss how \oursys{} partitions DNN operators into compute-shift-based execution plans.

\vspace{0.1em}
\noindent
\textbf{Operator representation.}
To represent an operator's computation, \oursys{} uses tensor expression \hlRA{\mbox{\cite{tensor_comprehensions, Halide, tvm, ansor, roller}}}, which defines how each output tensor value is computed from the input values. For example, a matrix multiplication of tensors $A$ in shape $[M,K]$ and $B$ in $[K,N]$ into $C$ is defined as
% \oursys{} uses tensor expression~\cite{tvm, ansor, roller} to represent an operator's computation. A tensor expression defines how each output tensor element is computed based on the input tensor values. For instance, a matrix multiplication of tensors $A$ in shape $[M, K]$ and $B$ in $[K, N]$ into $C$ can be expressed as
\vspace{-0.1em}
\begin{equation}\label{eqn:matmul_texpr}
    C[m, n] \mathrel{+}= A[m, k] * B[k, n],
    \vspace{-0.1em}
\end{equation}
where $m$, $k$, and $n$ are axes to index the elements in each tensor. \Cref{eqn:matmul_texpr} indicates that any value in $C$ indexed by $m$ and $n$ (i.e., $C[m, n]$) is computed by summing $A[m, k]*B[k, n]$ over all possible indices $k$.
\hlF{\oursys{} supports all common operators, like MatMul and Convolution,
from DNN workloads in both inference and training.}
For a few special cases like Sort, which cannot be represented in tensor expression,
\oursys{} uses the implementations from the vendor library.
% \oursys{} relies on the vendor library to provide their kernel implementations. 

\vspace{0.2em}
\noindent
\textbf{Partitioning an operator.}
To map an operator to interconnected cores, \oursys{} first partitions it into parallel \textbf{\textit{sub-operators}} along all unique axes in its tensor expression, using an \textit{operator partition factor} ($\bm{F_{op}}$).
% For example, the operator in~\Cref{eqn:matmul_texpr} can be partitioned along the $m$, $n$, and $k$ axes. We use a vector $F_{op}$ to indicate the partition factor for each corresponding axis of this operator. 
For example, \Cref{eqn:matmul_texpr} contains axes $m$, $k$, and $n$, then $F_{op}$ is a vector of three integer factors specifying how the three axes are spatially partitioned.
The total number of sub-operators is the product of all elements in $F_{op}$.
For example, $F_{op} = [2,1,3]$ for $[m,k,n]$ slices the operator into 6 sub-operators on 6 cores, each computing a $[{\frac{M}{2}}, {\frac{K}{1}}] \times [{\frac{K}{1}}, \frac{N}{3}]$ sub-matrix multiplication.

\vspace{0.1em}
\noindent
\textbf{Partitioning \rtensor{}s.}
% Based on the operator partitioning, we can determine the partitioning for all related tensors and transform them into \rtensor{}s.
\oursys{} then uses $F_{op}$ to derive the spatial partition factor $f_s$ for each tensor, following the data dependencies in tensor expression. 
% First, \oursys{} uses $F_{op}$ to derive the spatial partition factor $f_s$ for each tensor by analyzing the data dependencies in the tensor expression. 
% For example, the above $F_{op}=[3, 2, 1]$ implies the spatial factors $[3, 1]$, $[1, 2]$, and $[3, 2]$ for tensors A, B, and C. They are split into sub-tensors of shapes $[{M\over3}, K]$, $[K, {N\over2}]$, and $[{M\over3}, {N\over2}]$.
With the same example, for $F_{op} = [2, 1, 3]$ on $[m, k, n]$, the spatial partition factor for the tensor A is $f_s^A = [2, 1]$ for axes $m$ and $k$. Similarly, for tensors B and C, we have $f_s^B = [1, 3]$ and $f_s^C = [2, 3]$.
% If a tensor dimension is a combination of multiple axes, e.g., the third dimension of tensor $I$ in~\Cref{eqn:conv_texpr} is $h+k$, \oursys{} will partition 
% along this dimension to ensure each sub-operator has all the required data.
%with an overlapped region to ensure each sub-operator has all the required data. %, which is also achieved through expression analysis.

\begin{figure*}[t]
    \centering
    % \vspace{-3.1ex} % hotCRP compatible, do not change: -3.1ex scale=0.6
    \includegraphics[scale=0.6]{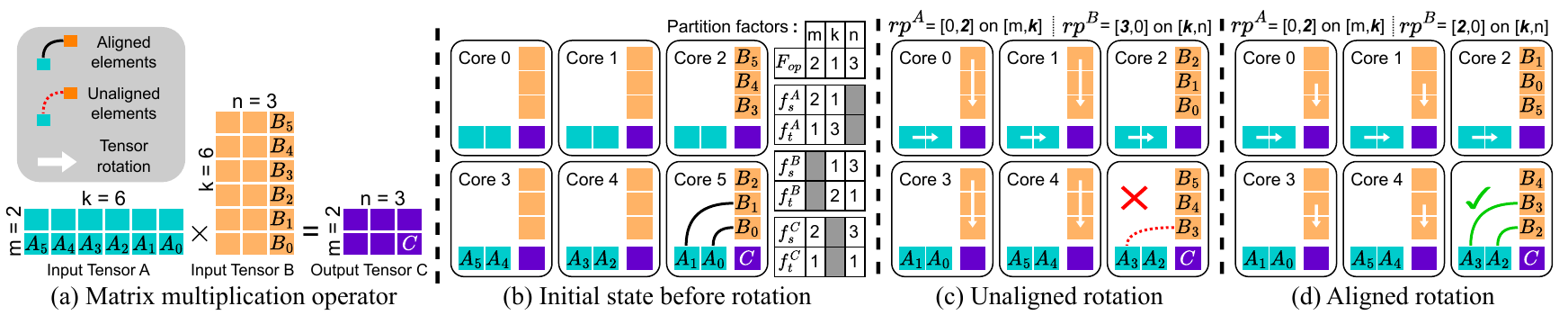}
    \vspace{-5ex}
    \caption{An example of the rotation of \rtensor{}. The compute-shift executions of the sub-operators need to be aligned.}
    \label{fig:cannon}
    \vspace{-1.5ex}
\end{figure*}

If a tensor's dimensions do not include some axis in $F_{op}$, each of the sliced sub-tensors is required by multiple sub-operators along the missing axis.
Thus, once the spatial factor determines the number of cores that will share a sub-tensor, the temporal factor determines how we split the sub-tensor across these cores into rotation ring(s).
In the above example, $F_{op}$ partitions the entire operator onto $2\times1\times3=6$ cores, and $f_s^B$ spatially partitions tensor B into $1\times3=3$ sub-tensors. Thus, each sub-tensor is shared by $P=\frac{6}{3}=2$ cores.
Then, a temporal factor $f_t^B=[2, 1]$ further splits each sub-tensor into $2\times 1 = 2$ partitions, forming $\frac{P}{2}=1$ rotation ring.

\pname{} enforces that the product of elements in $f_t$, or $\prod{f_t}$, is a divisor of the number of cores that shares the sub-tensor ($P$), so that the number of rotation rings (i.e., $\frac{P}{\prod{f_t}}$) is an integer.
If there is more than one rotation ring, we replicate each sub-tensor $\frac{P}{\prod{f_t}}$ times to ensure that each 
% sub-operator can access one partition at any given time.}
ring shares one copy of the sub-tensor.
While the duplication consumes memory space, it may reduce the number of rotation steps by allowing a larger sub-task on each core at each step, which enables a trade-off between memory usage and communication cost.

% For example, in Figure~\ref{fig:motivation_example} (b), the operator is partitioned by $F_{op} = [2, 1, 1]$ and the weight tensor (does not contain axis $m$) is required by both cores.
% \oursys{} can specify the temporal partition factor $f_t$ for this \rtensor{} and a rotating pace $rp$ to store and rotate these sub-tensors. 
% For example, by specifying $f_t=[1, 2]$ and $rp=[0, 1]$ for the weight \rtensor{}, we can map it to the two cores and rotate the sub-tensors for two rounds as shown in Figure~\ref{fig:motivation_example} (c).
% Note that, if a reduction dimension is partitioned, each processor will hold a partial result of the output tensor, which requires an additional reduction stage to obtain the full result.

\vspace{0.1em}
\noindent
\textbf{Aligning the rotations of \rtensor{}s.}
Since a general DNN operator can have various tensor shapes, a naive partitioning plan can easily cause the tensor shifting and the sub-task computing at an unaligned pace.
In \mbox{Figure~\ref{fig:cannon}} (a), we still use the MatMul operator in \mbox{\Cref{eqn:matmul_texpr}} as an example.
We partition it into a $2\times3$ grid in \mbox{Figure~\ref{fig:cannon}} (b), with the specified partition factors. Note that both A and B are temporally partitioned along axis $k$, but with different $f_t$ factors.

The rotating paces of tensors in one operator must be aligned to ensure correct data dependency.
In \mbox{Figure~\ref{fig:cannon}} (c), tensors A and B are shifted with different $\rp$s along axis $k$, which breaks data dependency.
In the bottom-right core, we cannot compute $C \mathrel{+}= A_2 * B_2$, as $B_2$ is not on this core after an unaligned rotation.
% By default, tensor A will be shifted in 2 steps and tensor B in 1 step, so sub-task is unable to access all its required data timely (Figure~\ref{fig:cannon} (c)).
% To organize computation into an aligned compute-shift plan,
Thus, \oursys{} synchronizes the $\rp$ of each \rtensor{} in Figure~\ref{fig:cannon} (d).
With $\rp=2$ on axis $k$ for tensors A and B, in each step, each core shifts A and B for 2 data elements along $k$, and computes a sub-task whose length along $k$ is also 2 (i.e., a sub-MatMul of shape [m=1, k=2, n=1]), requiring $\frac{6}{2}=3$ steps to finish the sub-operator on this core.

% \todo{equation 2 is not needed in reality; reconfirming}

To organize the computation into an aligned compute-shift plan, \oursys{} enforces two constraints.
First, if a set of \rtensor{}s rotate along the same axis $k$, they must share the same rotating pace $\rp$ along $k$.
% Second, the shared $\rp$ value on $k$ must be a divisor of the (spatially partitioned) sub-operator's length along $k$ (i.e., the entire operator's length along $k$ divided by its $F_{op}$ on $k$, which is $\frac{6}{1}=6$ for a sub-operator in \mbox{\Cref{fig:cannon}}), so we can complete the entire sub-operator in an integer number of steps.
Second, for each \rtensor{}, the $\rp$ value cannot exceed the length of its sub-tensor partition on dimension $k$, so that each $\rp$-aligned sub-task can be executed on the sub-tensor partitions locally on each core.
% Second, for each \rtensor{}, the $\rp$ must not be greater than the dimension length along $k$ of its sub-tensor partition on each core, so that each $\rp$-aligned sub-task can obtain a sub-tensor partition that is large enough to execute from the local memory of its corresponding core. 
% so that the data tile in each core's local memory is large enough to fulfill each $\rp$-aligned sub-task.
% Second, for each \rtensor{}, the $\rp$ along $k$ must not be greater than its sub-tensor partition dimension along $k$, so that all the required data for each sub-task is located in the local memory of each core.
As shown in \mbox{\Cref{fig:cannon}}, tensor A and B are partitioned by their $f_t$ along $k$ into dimension lengths of $\frac{6}{3}=2$ and $\frac{6}{2}=3$, respectively, so their $\rp$ on $k$ should not be greater than 2. 
To maximize compute intensity,
% \oursys{} sets $\rp$ to the minimum dimension length along the rotating axis $k$ of any sub-tensor partitions from the \rtensor{} set.}
\oursys{} designates the $\rp$ as the minimum of the sub-tensor partition lengths. 

With the above constraints, we can organize an operator's computation into a valid compute-shift execution plan. At each step, each sub-operator computes a sub-task partitioned by $F_{op}$ and the rotating pace $\rp$ along each axis. Each sub-operator iterates over all its sub-tasks by nested-looping through the axes of this operator.
Between sub-tasks, an \rtensor{} is rotated along the currently iterating axis for all its sub-tensors, until all sub-tasks are enumerated.
% When the computation is finished, all the \rtensor{}s complete a full rotation loop.
% Figure~\ref{fig:cannon} (d) illustrates an aligned MatMul operator execution plan on a $2\times3$ cores, which is partitioned by $F_{op}=[2, 3, 1]$. The corresponding \rtensor{}s are partitioned by A=\{$f_s:[2,1], f_t:[1, 3], rp:[0, K/3]$\}, B=\{$f_s:[1,3], f_t:[2, 1], rp:[K/3, 0]$\}, C=\{$f_s:[2,3], f_t:[1, 1], rp:[0, 0]$\}.

% \vspace{-3ex}
\subsection{Intra-operator and Inter-operator Trade-off}
\label{sec:design:optimization}
% \vspace{-0.3em}
% \paragraph{Intra- and inter-operator optimization space.}

For each operator, there could be a vast number of execution plans involving different spatial and temporal partition factors and rotating paces. 
%Each configuration may have different impacts on computing performance, communication cost, and memory footprint. 
Moreover, an end-to-end model consists of numerous operators, creating a substantial combinatorial optimization space.
% An unoptimized execution plan not only introduces significant overhead in computation and communication, but also wastes precious memory space.
% limiting the ability to fit larger models into the processor chip. 
% For each operator, there could be a vast number of potential execution plans involving different spatial or temporal partitioning factors, rotating speeds, and others. Each configuration may have different impacts on computing performance, communication cost, and memory footprint. Moreover, an end-to-end model consists of numerous operators, creating a substantial combinatorial optimization space.
% An unoptimized execution plan not only introduces significant overhead in computation and communication, but also wastes precious memory space, 
% limiting the ability to fit larger models into an intelligence processor. 
% To optimize the execution time with the best memory space utilization, 
\oursys{} defines a \textit{two-level trade-off space between execution time and memory consumption}.

First, when determining each operator's execution plan, we can trade memory space for execution efficiency by specifying a smaller temporal partitioning factor.
This can reduce communication costs by reducing the hops in the rotation loop, while at the cost of using more memory to hold duplicated tensors.
We refer to this as \textbf{\textit{intra-operator trade-off}}.

Second, we can tradeoff between memory space and execution time across all operators holistically when deciding the end-to-end model execution plan.
Different operators have different memory-latency trade-offs, allowing us to allocate more memory to operators with higher memory-cost efficiency.
Moreover, a single operator can have multiple execution plans, such as utilizing a memory-efficient plan when the operator is not executing to save more memory space, and switching to an execution-efficient plan when it is about to execute.
We refer to this as \textbf{\textit{inter-operator trade-off}}.

To optimize such two-level trade-offs, \oursys{} decouples the combinatorial space into two distinct stages.
First, \oursys{} optimizes each individual operator's execution by searching for all Pareto-optimal trade-off plans between execution time and memory consumption ($\S$\ref{subsubsec:intra-operator}).
Second, \oursys{} globally optimizes the memory allocation among different operators based on the intra-operator Pareto-optimal plans ($\S$\ref{subsubsec:inter-operator}).

% \noindent
\subsubsection{Searching Pareto-optimal Intra-operator Plans}
\label{subsubsec:intra-operator}
% \vspace{-0.5em}

For each operator, to search for the optimal execution plan among numerous configuration choices, an efficient performance feedback for each plan is essential. \oursys{} leverages the unique advantage of distributed on-chip memory architecture, which involves only local computation and communication, to design a sufficiently accurate cost model.

First, given the partitioning factors of an operator and its tensors, for each plan we can statically derive performance factors such as per-core computation task, per-step communication volume, and memory footprint.
Second, the computation on each core in each step only consumes data from local memory, preventing unpredictable memory stalls from lower-level memory.
\hlD{To build a cost model for each operator type, we randomly generate sub-tasks with different shapes, run them on a single IPU core, and profile their execution time. We fit a linear regression model using the sub-task shape as input and the execution time as output.}
% For each operator type, we fit a linear regression model by taking factors such as total FLOPS, tensor shape, and number of steps as inputs to predict the execution latency.
\hlD{Also, an interface is exposed for users to implement custom cost functions
for their custom kernels.}
In addition, the communication time is also accurately fitted by a linear regression model that takes the data transfer volume as input.

With the cost model, \oursys{} quickly examines each execution plan and chooses the best candidates that sit on the Pareto optimal trade-off curve, where each plan either runs faster than any other plans with the same or less memory footprint or uses less memory than any others with the same or lower execution time (see Figure~\ref{fig:intra_performance} in our evaluation for details). 

\begin{figure}[t]
    \centering
    \includegraphics[width=1\linewidth]{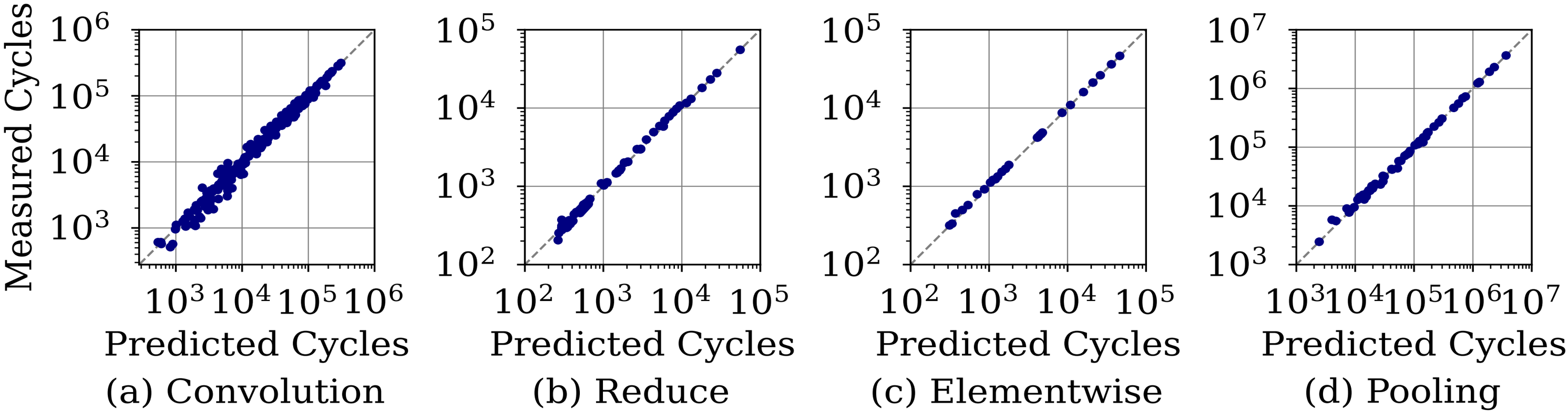}
    \vspace{-5ex}
    \caption{Cost model accuracy for different operator types and shapes. Each point represents the measured vs. predicted execution time of a sub-operator.}
    \label{fig:intra_cost}
    \vspace{-2.5ex}
\end{figure}

% \vspace{0.2em}
\noindent
\textbf{Cost model accuracy.}
To test the accuracy of the cost models, we vary the operator shape and compare the predicted execution time with the actual profiled execution time. \Cref{fig:intra_cost} shows the accuracy of representative operator types. For most operators, \pname{} achieves near-perfect accuracy. The only exception is convolution, which is implemented with vendor-supplied kernels that apply some black-box optimizations. Even with slight inaccuracy, \pname{} can still find sufficiently good execution plans and outperform state-of-the-art compilers (see $\S$\ref{sec:eval_intra_op}). We envision that hardware vendors would be able to supply a perfect cost model for their kernels as they integrate \pname{} in their toolchain.

% \vspace{0.2em}
% \noindent
% \textbf{Search policy.}
% [TODO]

% \vspace{0.2em}
\noindent
\textbf{Search constraints.}
When \pname{} enumerates all possible execution plans,
a large portion of plans are evidently inefficient, and it is unnecessary to evaluate them with the cost model.
Thus, \oursys{} employs a rule-based approach to filter out inefficient plans, by applying two user-configurable constraints.

First, a plan utilizing too few cores will cause compute underutilization.
Thus, the \textit{parallelism constraint} specifies the minimum number of cores an operator uses, filtering out plans with low parallelism.
\hlRA{For example, 
% the number of possible values of each element in $F_{op}$ is bounded from above by the minimum of (1) the number of cores and (2) the length of each dimension, as $F_{op}$ specifies the number of spatial partitions along each dimension.
to run a 1-dimension operator with dimension length $L$ on $C$ cores, we may partition it into 1 to $\min(L, C)$ sub-operators, which utilizes 1 to $\min(L, C)$ cores.
Thus, there are $\min(L, C)$ possible values for $F_{op}$.
To utilize at least 90\% of the cores, we must partition it into $0.9$$\times$$\min(L, C)$ to $\min(L, C)$ sub-operators, so we only need to enumerate $\frac{\min(L, C)}{10}$ possible values for $F_{op}$.}

Second, to leverage the matrix accelerator unit (e.g., AMP in IPU~\mbox{\cite{ipu_hardware}}) in each core, we may need to pad the tensor shape to align with hardware, which underutilizes the memory capacity and FLOPS.
Thus, the \textit{padding constraint} specifies the maximum padding as the ratio between the original tensor size and the padded tensor size, filtering out plans with excessive padding.
\hlRA{For example, when partitioning a dimension with length $L$ into $p$ partitions with length $l$, we calculate the ratio between the original length and padded length
% original v.s. padded ratio 
as $\frac{L}{l p}$. Plans with ratios below a certain threshold (e.g., a threshold of 0.9 means that the max padding overhead is $\frac{1}{0.9}-1=11\%$) will be discarded.}

% \hlC{Both constraints are set to 0.9 by default.}
% This can introduce significant waste if the padding size is too large.
% A large padding can significantly waste memory space and FLOPS.
% To avoid inefficient padding, \oursys{} employs a rule-based approach~\cite{roller} to filter out inefficient plans. %, which typically constitute the majority of the space.
Only the remaining plans are evaluated by the cost model.
% We apply two user-configurable constraints to filter inefficient execution plans in $\S$\ref{sec:design:optimization}:
% (1) the \textit{parallelism constraint} specifies the minimum number of cores used by an operator, which filters out plans with low parallelism;
% % (2) the \texttt{compute padding constraint} specifies the maximum padding size of the AMP unit in each core;
% and (2) the \textit{padding constraint} specifies the maximum padding as the ratio between the original tensor size and the padded tensor size, which filters out plans with large padding sizes that waste both memory and FLOPS.
We examine the impact of both constraints in
\S\ref{sec:eval_intra_op}.
% We quantify the search space size reduction in Figure~\ref{fig:intra_search}. 

\begin{figure}
    \centering
    % \vspace{-3ex} % hotCRP compatible, do not change:
    \includegraphics[width=1\linewidth]{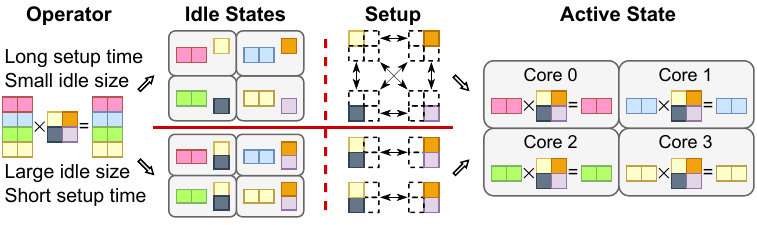}
    \vspace{-5ex}
    \caption{Switching operator state from idle to active. There is a tradeoff between the memory footprint of idle states vs. the time of turning idle states to active states (setup time). }% Deduplicate idle operators
    \label{fig:idle}
    \vspace{-2ex}
\end{figure}

\begin{figure*}
    \centering
    % \vspace{-0.5ex} % hotCRP compatible, do not change: -3.7ex width=0.88\linewidth
    \includegraphics[width=1\linewidth]{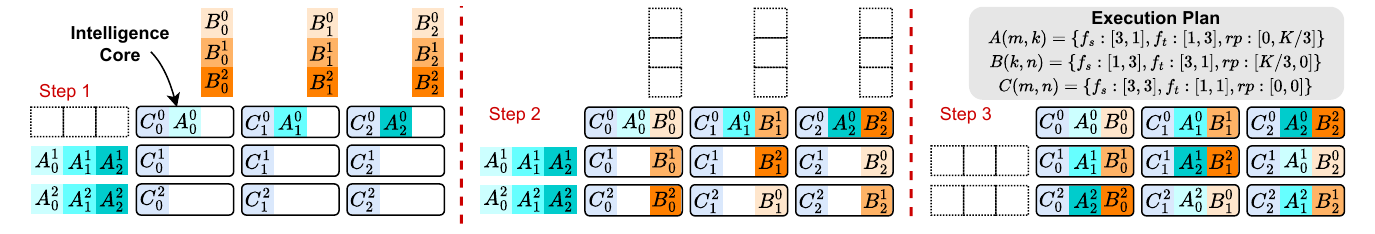}
    \vspace{-5.5ex}
    \caption{Sub-tensor placement for a matrix multiplication operator on 3$\times$3 cores.}
    \label{fig:placement}
    \vspace{-1ex}
\end{figure*}

% \vspace{0.2em}
% \noindent
% \vspace{-2.5ex}
\subsubsection{Holistic Inter-operator Memory Reconciliation}
\label{subsubsec:inter-operator}
% \vspace{-0.3em}

% \newlength{\oldtextfloatsep}
% \setlength{\oldtextfloatsep}{\textfloatsep}
% \setlength{\textfloatsep}{0pt}% Remove \textfloatsep

To execute a DNN model, \oursys{} adopts the common approach of fitting multiple operators into on-chip memory, 
while only transferring the input and output data through off-chip memory.
This strategy is advantageous as adjacent operators can reuse full intermediate data in the fast on-chip memory.

Given the limited on-chip memory, each operator in \oursys{} is assigned two execution plans — one for minimum memory usage before operator execution (called \textit{idle state}) and one for minimum latency during execution (called \textit{active state}). When an operator begins to execute, a \textit{plan setup phase} transforms the partitioning plan from idle state to active state, i.e., by transferring the necessary data through the inter-core connection. 
This creates a tradeoff between memory usage at idle state and the overhead at the setup phase.
Figure~\ref{fig:idle} shows the state transition of a MatMul operator.

As each operator has many choices for idle and active plans, \oursys{} trades-off globally with a holistic inter-operator reconciliation policy, as shown in Algorithm~\ref{alg:inter-op}.
Initially, \oursys{} assigns the memory-efficient plan as the idle plan for all operators (line 3).
The remaining memory space after placing all operators is considered as the \textit{active memory}\footnote{\textit{Active memory} is the memory space used by the operator currently in active state, and \textit{idle memory} is the space used by operators in the idle state.}, which is used for executing the sub-operators.
\oursys{} then searches for the best execution plan by greedily trading more active memory space for \textit{idle memory} space, aiming to reduce more setup time at the cost of slightly increasing execution time.

\newcommand\mycommfont[1]{\footnotesize\sffamily\textcolor{purple}{#1}}
\SetCommentSty{mycommfont}
% \SetAlFnt{\small\sffamily}
% \newlength{\textfloatsepsave}
% \setlength{\textfloatsepsave}{\textfloatsep}
% \setlength{\textfloatsep}{0pt}
% \begin{figure}[t]

\begin{algorithm}[t]
% \footnotesize
\small
\DontPrintSemicolon
\SetAlgoNoEnd
\SetAlgoLined
% \SetAlgoHangIndent{0em}
% \Setvlineskip{-4ex}
% \SetAlgoSkip{0ex}
% \KwIn{{\it all\_ops} $\gets$ }
% \SetKwBlock{Begin}{Function}{End Function}
\SetKwFunction{FMain}{{\it get\_best\_idle\_configs}}
\SetKwProg{Fn}{Function}{:}{}
% \Begin(get\_best\_idle\_configs([[op0\_configs], [op1\_configs], ...]))
% \Fn{\FMain{[[op0\_configs], [op1\_configs], ...]}}
\Fn{\FMain{all\_ops}}
{
  \tcp{\color{purple}{start from the memory-efficient plan}}
  \For{op in all\_ops}
  {
    \textit{idle\_plan[op]} $\gets$ plan with min memory use \;
  }
  \textit{idle\_mem\_size} $\gets$ $\sum_{\textit{op} \in \textit{all\_ops}}\textit{idle\_plan[op].size}$\;
  \textit{cur\_best\_time} $\gets$ $\infty$ \;
  \While{idle\_mem\_size < max\_mem\_per\_core}
  {
    \tcp{\color{purple}{update active plan for each op}}
    \For{op in all\_ops}
    {
      \textit{active\_plan[op]} $\gets$ fastest plan that fits in mem \;
    }
    % \tcp{\color{purple}\textbf{evaluate current plan}}
    \textit{time} $\gets$ \textit{estimate\_total\_time(idle\_plan, active\_plan)} \;
    \If{time < cur\_best\_time}
    {
      % \tcp{\color{purple}\textbf{update the current best plan}}
      \textit{cur\_best\_time} $\gets$ \textit{time} \;
      \textit{cur\_best\_plan} $\gets$ \textit{(idle\_plan, active\_plan)} \;
    }
    \tcp{\color{purple}{find the best operator and increase its idle mem size}}
    \textit{best\_op} $\gets$ op with highest $-\Delta T_S /\Delta M_I$ \;
    update \textit{idle\_plan[best\_op]} with new plan \;
    update \textit{idle\_mem\_size} based on the new \textit{idle\_plan} \;
  }
  \Return{cur\_best\_plan}
}
\caption{Inter-op memory reconciliation policy.}\label{alg:inter-op}
\end{algorithm}

For each search step, \oursys{} first selects the best operator,
which has another execution plan that reduces the most setup time while adding the smallest idle space.
This plan is found by computing the ratio $\Delta T_S /\Delta M_I$ for each operator, where $\Delta T_S$ and $\Delta M_I$ are the reduced setup time and the increased idle space
(line 13).
To apply the plan change, the active memory space is subtracted by $\Delta M_I$
% After selecting the best plan, the active space is updated by subtracting $\Delta M_I$.
(line 15).
\oursys{} updates the total execution time by finding the fastest execution plan that fits in the given active memory for each operator (line 8),
and adding up the latencies of all operators (line 9).

The inter-operator scheduling policy of \oursys{} can explore the complex search space with low algorithmic complexity. While there are $\prod_{i=0}^{\# ops}(op[i].num\_idle\_plans)$ idle plan combinations in total, we acquire an optimized one by searching only $\sum_{i=0}^{\# ops}(op[i].num\_idle\_plans)$ promising combinations. Given the moderate number of possible trade-off configurations, \oursys{} currently searches all steps and chooses the one that can lead to the minimum end-to-end execution time.

% \begin{algorithm}[t]
% \footnotesize
% \caption{Inter-operator memory reconciliation policy.}
% \label{alg:priority}
% \begin{algorithmic}[1]
% \Function{get\_best\_idle\_configs}{[[op0\_configs], [op1\_configs], ...]}
%     \For{op in all\_ops}
%         \State idle\_plan[op] <- plan with min memory use
%     \EndFor
%     \While{sum(idle\_plan[].mem) < max\_mem\_per\_core}
%         \For{op in all ops}
%             \State exec\_plan[op] <- fastest plan fits in mem
%         \EndFor
%         \State time <- estimate\_total\_time(idle\_plan, exec\_plan)
%         \If{time < cur\_best\_time}
%             \State cur\_best\_time <- time
%             \State cur\_best\_plan <- (idle\_plan, exec\_plan)
%         \EndIf
%         \State increment idle plan for op with highest $-\Delta T_W /\Delta M_I$
%     \EndWhile
%     \State return cur\_best\_plan
% \EndFunction
% \end{algorithmic}
% \end{algorithm}

% \vspace{-2ex}
\subsection{Mapping to the Hardware Accelerator}
\label{sec:design:hardware}
% \vspace{-0.5em}
% \setlength{\textfloatsep}{\textfloatsepsave}

The compilation approach of \oursys{} is designed to be extensible for general distributed on-chip memory-based accelerators, which can be abstracted into a unified architecture with multiple cores, each equipped with dedicated local memory and interconnected via a high-speed on-chip network.

% \vspace{0.2em}
\noindent
\textbf{Abstracted device interface.}
% To facilitate adaptation to different accelerators, 
\oursys{} abstracts three key device interfaces: 
% \texttt{allocate}, \texttt{compute}, and \texttt{shift}
\textbf{(1)} \texttt{allocate} serves as a compile-time interface to allocate memory space for placing tensor partitions.
\textbf{(2)} \texttt{compute} functions as a code generation interface that emits instructions for computing a specific sub-operator on a core. By default, \oursys{} utilizes a few pre-defined code templates to generate single-core computing logic for each specific partition configuration. 
\textbf{(3)} \texttt{shift} serves as a runtime communication primitive to transmit a sub-tensor to the specified destination core.
\oursys{} leverages these interfaces to map the optimized execution plan to an accelerator.

% \vspace{0.2em}
\noindent
\textbf{Sub-tensor placement.}
\oursys{} allocates the entire memory space and assigns each sub-tensor to its corresponding core. To optimize memory consumption, \oursys{} performs tensor liveness analysis to reuse the memory of precedent operators. To ensure that sub-tensors from different tensors are in the same core at each rotating step, \oursys{} arranges the initial placement of each tensor partition step-by-step by analyzing the computing order of each sub-operator and their data dependencies.
This ensures that (1) the initial placement of all sub-tensor partitions satisfies the data dependency on each core, and (2) sub-tensor partitions along each axis are in ascending order, guaranteeing that the data dependency on each core is still satisfied after each rotating step.

We show an example of placing sub-tensors for a 3$\times$3 matrix multiplication (i.e., \Cref{eqn:matmul_texpr} with $M=K=N=3$) in \Cref{fig:placement}.
Initially, all output sub-tensors ($C_j^i$) are allocated based on the partitioning plan.
Then, the first sub-tensor set $\{A^0_0, A^0_1, A^0_2\}$ is partitioned along the temporal dimension and distributed to the corresponding cores (the first row of cores).
Given the current placement of $A^0_i$ and $C^0_i$, we infer that $B_i^i$ should be placed in the first row of cores due to data dependency.
Subsequently, the remaining $B_i^j$ sub-tensors are placed sequentially in each column of cores.
Following this process, the placement of all $A_i^j$ sub-tensors can be inferred.

\noindent
\textbf{Sub-operator computation scheduling.}
After the tensor placement, \oursys{} organizes the computation on each core as nested loops of interleaved \texttt{compute} and \texttt{shift} stages. Each loop corresponds to a temporal partition number.
% For example, if there are two dimensions to shift, $m$ and $n$, with $m$ shifting twice and $n$ shifting once, then there are two valid 5-time shift schedules: (1) $m$ is the inner loop, i.e., shift($m$)$\times 2\to$ shift($n$) $\to$ shift($m$)$\times 2$; and (2) $n$ is the inner loop, i.e., shift($n$) $\to$ shift($m$) $\to$ shift($n$) $\to$ shift($m$) $\to$ shift($n$).
To determine the optimal loop order, \oursys{} designates the dimension belonging to the tensor with the smaller size as the inner loop to reduce the total communication volume, as the inner loop is executed more times.
% During shifts, each core performs one round of computation on its local data.
To generate local computations for each core, \oursys{} invokes the corresponding \texttt{compute} function with the partition configuration and the tensor expression.

%% file: implement.tex
% \vspace{-0.8em}
\section{Implementation Details}
\label{sec:implementation}
% \vspace{-0.6em}

%\pname{} is implemented in Python and it 
%\pname{} leverages NNFusion as the front-end to process the input DNN models~\cite{nnfusion} and generate a sequence of tensor operators. 
We implement \pname{} as a standalone compiler framework and adapt it to Graphcore IPU,
% \footnote{IPU is the only distributed on-chip memory accelerator we can access.}
a representative accelerator with distributed on-chip memory.  
\pname{} takes a DNN model in ONNX format~\cite{onnx} as input and parses it into an operator graph, where each operator is represented by a tensor expression.
It performs the intra- and inter-operator optimizations, and outputs the executable kernel code for IPU.
The optimization passes are implemented with 4K LoC of Python.
The kernel code generation is developed with 4.5K LoC of Python and 1.5K LoC of C++ and IPU assembly~\cite{ipu_isa}.

% \subsection{Kernel Code Generation}
% \label{sec:implem:codegen}

\begin{figure}[t]
    \centering
    % \vspace{0.5ex}
    \includegraphics[width=1\linewidth]{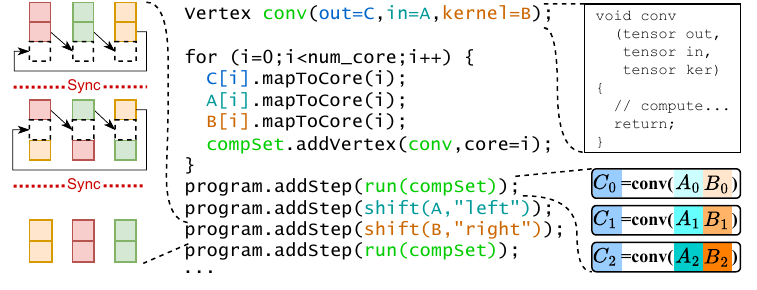}
    \vspace{-5ex}
    \caption{The kernel code example of \pname{} on IPU.}
    \label{fig:codegen}
    \vspace{-2ex}
\end{figure}

% \vspace{0.2em}
\noindent
\textbf{Kernel code generation.}
\Cref{fig:codegen} shows an example of the code generated for IPU. %, a real world instance of DSM hardware. 
First, \pname{} \texttt{allocate}s the tensors using the \texttt{t.mapToCore(int i)} interface, which maps the sub-tensor \texttt{t} to core \texttt{i}.
% For \texttt{compute} operations, \pname{} assigns per-step local tasks, called \texttt{Vertices} on IPU, to each core.
And then, \pname{} uses the \texttt{compute} interface to generate the per-step local computation task on each core, which is called \texttt{Vertex} on IPU.
In each step, each core runs its own \texttt{Vertex}.
% of the current operator for once using the data currently inside its on-chip memory.
%, where each sub-operator is call a \texttt{Vertex} in IPU.
% A \texttt{Vertex} is assigned to one specific core, while 
All \texttt{Vertices} executed in the same step are homogeneous in \pname{} and form a \texttt{ComputeSet}. 
% Thus all cores in the IPU chip execute one \texttt{ComputeSet} per round.
\pname{} schedules data \texttt{shift}s across all cores.
% Between each step, \pname{} schedules \texttt{shift}s on all cores to shuffle data across cores.
% \pname{} guarantees the correctness of computation by correctly placing and shifting the data among cores.
After a \texttt{shift}, all cores can execute the same \texttt{ComputeSet} with the new local data.
Once an operator finishes, the program may rearrange the data on cores if necessary, before launching the next one.
\noindent
\textbf{Multi-copy shift with buffer.}
% While \pname{} relies on the \texttt{shift} interface,
The IPU does not support the \texttt{shift} interface by default. % as it is not a true dataflow architecture.
% While shift is data movement whose source space is overlapped with destination space, the trivial data copy is data movement that has no overlap between source and destination. 
The key limitation is that the source (data being shifted out) and the destination (data being shifted in) overlap in memory.
To overcome this issue, we implement a low-overhead pseudo-shift mechanism as shown in the left side of \Cref{fig:codegen}. We use a temporary buffer in each core, as shown by the dashed boxes, to avoid overlapping source and destination.
% We observe two overheads of this mechanism. First, the temporary buffer occupies on-chip memory space, so it cannot be too large. Second, if the buffer is too small, the shift will require multiple iterations.
A larger temporary buffer consumes more on-chip memory space, while a smaller buffer will require multiple shift iterations. 
\pname{} reserves an 8KB buffer in each core's local memory by default, 
which incurs negligible synchronization overhead,
while allowing users to configure the buffer size by themselves.
% We tune the buffer size for each hardware device separately to find the optimal setting.
% On IPU Mk2, a 118-cycle (89-ns) synchronization occurs for each shift iteration, while the data transfer rate per inter-core link is 3.57 bytes/cycle. 
% \hl{To keep the synchronization overhead below a threshold (i.e., P\% of the entire inter-core transfer time), T10 requires} 
% \[
%     \texttt{buffer} = \texttt{sync\_overhead} * \texttt{inter\_core\_bandwidth} / \texttt{P\%}.
% \]
% For IPU, the buffer size for keeping the overhead below 5\% is $118\texttt{cycle} * 3.57\texttt{bytes\_per\_cycle} / 5\% = 8\texttt{KB}$.

%\pname{} allows the user to configure the buffer size. In practice, tuning the buffer size has negligible benefits on IPU, since the synchronization overhead is already very low.
%Thus, we keep the synchronization overhead below 5\% by statically reserving an 8KB buffer in each core's local memory.

% Thus, we keep the synchronization overhead below 5\% by statically reserving an 8KB buffer in each core's 624KB on-chip memory, which occupies a negligible 1.28\% of memory. 
% Each data transfer round takes at most $\frac{8\times1024\text{ B}}{3.57\text{ B}/\text{cycle}} + 118 \text{ cycles} = 2413 \text{ cycles}$, where only 4.9\% of the cycles are synchronization overhead.

% \vspace{0.2em}
\noindent
\textbf{Compound axis in tensor expressions.}
\pname{} supports all common tensor expressions in DNN models, including those with compound axes (i.e., an axis that is a function of other axes).
For example, a 2D Convolution can be expressed as
\vspace{-0.3em}
\begin{equation}
    O[b, f, h, w] \mathrel{+}= I[b, c, h+kh, w+kw] * C[f, c, kh, kw],
    \vspace{-0.3em}
\end{equation}
where $b$ is batch size, $c$/$f$ is the number of input/output channels, $h$/$w$ is output height/width, and $kh$/$kw$ is kernel height/width.
For the compound axes $h+kh$ and $w+kw$,
\pname{} partitions each basic axis (e.g., $h$ and $kh$) individually.
% we only partition the $h$ and $w$ dimensions, as convolutions usually have small kernels that are better duplicated across cores.
% We add paddings to these two dimensions in each sub-tensor to ensure that the data required for computation is included.

\noindent
\hlcommon{\textbf{Inter-operator transition.}
\pname{} allows two consecutive operators to have different tensor partitioning plans. If the input and output tensor layouts do not match, \pname{} inserts an all-to-all inter-core data exchange operation to adjust the layout. The overhead is typically small compared to the operator execution, since the intermediate tensor size is small compared to the inter-core shift volume during execution.}

% For the tensor $I$, its third dimension is a compound axis of $h$ and $kh$.
% If $h$ or $k$ is partitioned, we perform bound analysis on the compound axis $h+k$ to find the upper bound of the sub-tensor size~\cite{tvm}, and then we apply padding to the sub-tensor.
% If a tensor dimension is a combination of multiple axes, e.g., the third dimension of tensor $I$ in~\Cref{eqn:conv_texpr} is $h+k$, \oursys{} will partition 
% along this dimension to ensure each sub-operator has all the required data.
%with an overlapped region to ensure each sub-operator has all the required data. %, which is also achieved through expression analysis.

% \vspace{0.2em}
% \noindent
% \textbf{Intra-operator search constraints.}
% We apply two user-configurable constraints to filter inefficient execution plans in $\S$\ref{sec:design:optimization}:
% (1) the \textit{parallelism constraint} specifies the minimum number of cores used by an operator, which filters out plans with low parallelism;
% % (2) the \texttt{compute padding constraint} specifies the maximum padding size of the AMP unit in each core;
% and (2) the \textit{padding constraint} specifies the maximum padding as the ratio between the original tensor size and the padded tensor size, which filters out plans with large padding sizes that waste both memory and FLOPS.
% \hl{We examine the effects of both constraints in}
% \S\ref{sec:eval_intra_op}.

%% file: evaluation.tex
% \vspace{-0.5ex}
\section{Evaluation}
\label{sec:evaluation}
% \vspace{-0.5em}

We show that (1) \pname{} outperforms state-of-the-art DNN compilers on Graphcore IPU by up to 3.3$\times$ (1.69$\times$ on average)
% and enables individual operators to be up to 10.79$\times$ faster
($\S$\ref{sec:eval_e2e_perf});
(2) it enables flexible and efficient intra- and inter-operator scheduling ($\S$\ref{sec:eval_intra_op} and $\S$\ref{sec:eval_inter_op});
(3) it scales as we increase the number of cores ($\S$\ref{sec:eval_sens_vary_core_mem}); 
(4) it unleashes the benefit of distributed on-chip memory compared to a popular shared-memory-based AI chip - A100 GPU ($\S$\ref{sec:eval_gpu});
and (5) it benefits LLMs by alleviating the memory bandwidth wall ($\S$\ref{sec:eval_llm}).
% (5) \pname{} enables DSM-based accelerators to outperform the state-of-the-art GPU by up to 3.06$\times$ ($\S$\ref{sec:eval_gpu}).

\begin{table}[t]
    \centering
 %   \vspace{-0.5ex}
    \caption{DNN models used in our evaluation.}
    \vspace{-2.5ex}
    \scriptsize
    \begin{tabular}{|c|c|c|}
    \hline
        \textbf{Name} & \textbf{Description} & \textbf{\# of Parameters} \\\hline
        BERT~\cite{bert} & Natural Language Processing & 340M \\\hline
        ViT~\cite{vit} & Transformer-based Vision & 86M \\\hline
        ResNet~\cite{resnet} & CNN-based Vision & 11M \\\hline
        NeRF~\cite{nerf} & 3D Scene Synthesis & 24K\\\hline
        OPT~\cite{opt} & Large Language Model & 1.3B to 13B\\\hline
        Llama2~\cite{llama2} & Large Language Model & 7B to 13B\\\hline
        RetNet~\cite{retnet} & State Space Model for Language & 1.3B\\\hline
    \end{tabular}
    \label{tab:benchmarks}
    \vspace{-0.5ex}
\end{table}

\begin{figure*}[t]
    \centering
    \vspace{-0.5ex} % hotCRP compatible, do not change: -3.7ex scale=0.65
    \includegraphics[scale=0.68]{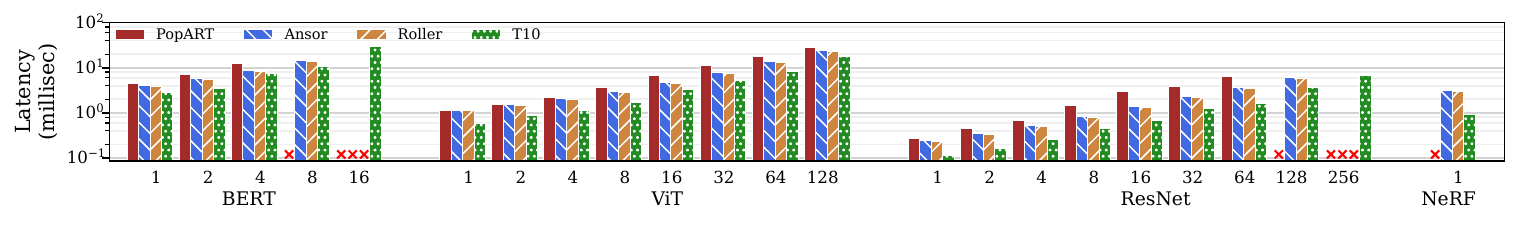}
    \vspace{-3.5ex}
    \caption{Inference latency of DNN models for various batch sizes. ``\textcolor{red}{\ding{54}}'' indicates the program cannot fit into an IPU chip.}
    \label{fig:e2e_performance}
    \vspace{-1.5ex}
\end{figure*}

\input{figures/table_compare}

% \vspace{-.5ex}
\subsection{Experimental Setup}
\label{sec:eval_setup}
% \vspace{-0.3em}

We evaluate \pname{} with DNNs of different types and sizes in \Cref{tab:benchmarks}, including CNNs (ResNet), Transformers (BERT and ViT), and fully-connected networks (NeRF). 
% \hl{Notably, intermediate tensors also occupy a significant amount of memory, and their sizes will grow as batch size increases.}
%We vary the batch size to demonstrate \pname{}'s ability to support larger models and batch sizes.
For each model, we test from batch size 1, and double the batch size until the program cannot fit into the on-chip memory. 
We also evaluate LLM decoding workloads using LLMs like OPT and Llama2 (see $\S$\ref{sec:eval_llm}). 
\hlF{Our evaluation focuses on inference, because the IPU chip we can access has limited on-chip capacity and is mostly used for inference.}

We execute models on a Graphcore IPU MK2 chip~\cite{ipu2} (see \Cref{tab:compare_arch}).
%The detailed hardware specifications are shown in~\Cref{tab:compare_arch}.
%We increase the batch size for each model until it cannot fit into the 900MB on-chip memory. 
We compare \pname{} with vendor libraries and current DL compilers that support distributed on-chip memory,
including Graphcore's official Poplar Advanced Run Time (\texttt{PopART})~\cite{popart} and two DL compilers based on VGM, \texttt{Ansor}~\cite{ansor} (modified to support IPU) and \texttt{Roller}~\cite{roller}.
As \pname{} only applies lossless optimizations without changing any arithmetic operations in a model,
the inference accuracies of T10, Roller, and PopART have negligible differences given the same pretrained parameters.

% \vspace{-2.5ex}
\subsection{End-to-End Performance}\label{sec:eval_e2e_perf}
% \vspace{-0.3em}

%We show the inference latency of each DNN model as we increase the batch size in \Cref{fig:e2e_performance}. 
As shown in \Cref{fig:e2e_performance}, \pname{} achieves 1.69$\times$ end-to-end inference latency improvement on average than Ansor and Roller. \pname{} supports larger batch sizes and models, while other baselines fail to execute as the batch size gets larger.
% \pname{} also enables IPU Mk2 to achieve better performance than the A100 GPU for some benchmarks, which is impossible for PopART and Roller.

% The baseline PopART selects the size of sub-operators (i.e., the work done on each core) based on some black box heuristics. It also divides the scratchpad memory in each core into a global cache region and a local buffer region. 
% This leads to two limitations.
% First, the global cache region always contains redundant data that are not accessed during compute phases. This limits the effective local buffer size and hurts the compute density of each sub-operator.
% Second, PopART incurs significant setup overhead as it performs inter-core data transfer to load the data from the global cache to each core's local buffer.
% Third, PopART generates sub-optimal sub-operators that cannot fully utilize the memory in each core. This leads to low compute throughput because frequent shift phases are needed to fetch data from other cores' global cache regions. As a result, PopART has the worst performance for all benchmark models.

% \vspace{0.2em}
\noindent
\textbf{DNN inference latency.}
Ansor and Roller can outperform PopART by 1.33$\times$ and 1.38$\times$ on average. They improve single-operator performance by using appropriate sub-operator sizes, such that each sub-operator utilizes more local memory, leading to higher data reuse and compute intensity. They have similar performance by exploring the same optimization space. 
However, they still suffer from significant inter-core communication overhead, due to the VGM abstraction. They also cannot make globally optimized decisions, as they only consider single-operator performance.

\pname{} outperforms Roller by up to 3.3$\times$ (1.69$\times$ on average).
% by exploring the trade-off between inter-core communication overhead and memory capacity overhead with both intra- and inter-operator scheduling.
%Compared to PopART and Roller, 
It avoids data duplication caused by VGM and efficiently utilizes on-chip memory with intra-operator scheduling, which enables larger sub-operator sizes and reduces the inter-core communication overhead.
With the holistic inter-operator scheduling, \pname{} reduces the setup and execution time of one operator with minimum negative impact on other operators.
% Its benefit is significant with small operators (e.g., ResNet with small batch sizes), as it eliminates most inter-core data transfers by setting up data in advance.
% as their low compute intensity cannot amortize the setup time with Roller.
% as their low compute intensity cannot amortize the virtual global memory load overhead in Roller.
%by eliminating unnecessary on-chip data duplication, exploring the complete spatial-temporal search space for each operator, and efficiently searching for a globally optimized DNN model execution plan. As a result, \pname{} .

% In addition to achieving better performance at the same batch size,
\pname{} also supports larger batch sizes. For example, PopART fails to execute the largest batch size of most models (except ViT) and cannot execute NeRF at all. Even though Roller can run a larger benchmark by selecting smaller sub-operators, it incurs much more performance penalty than \pname{}, %due to smaller sub-operator memory,
as it needs to fetch data frequently from the virtual global memory.
% In contrast, \pname{} reduces the on-chip data duplication with the intra-operator temporal mapping ($\S$\ref{sec:design:intra_op}) and can fit a larger program with a much smaller penalty.

% \vspace{0.2em}
% \noindent
% \textbf{Comparison with A100 GPU.}

% \begin{figure}[t]
%     \centering
%     \vspace{-2ex}
%     % [trim={left bottom right top},clip]
%     \includegraphics[width=0.9\linewidth,trim={0 0 0 2ex},clip]{figures/eval_breakdown.pdf}
%     \vspace{-3ex}
%     \caption{End-to-end performance breakdown of different DNN models with varying batch sizes on \pname{}.}
%     \label{fig:breakdown}
%     \vspace{-3ex}
% \end{figure}

\begin{figure}
    \centering
    % \vspace{-0.5ex}
    \includegraphics[scale=0.45]{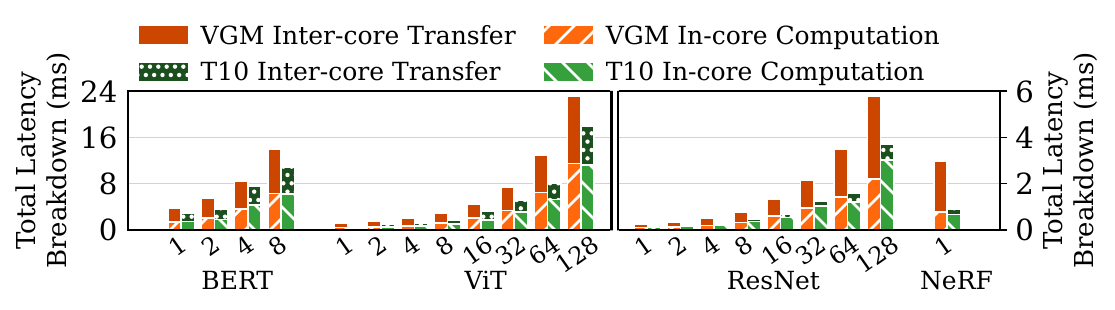}
    \vspace{-6.5ex}
    \caption{Data transfer overhead of executing various DNN models with different batch sizes on IPU.}
    \label{fig:compute_shift_breakdown}
    \vspace{-1.5ex}
\end{figure}

\begin{figure}
    \centering
    % \vspace{-0.5ex}
    \includegraphics[scale=0.48]{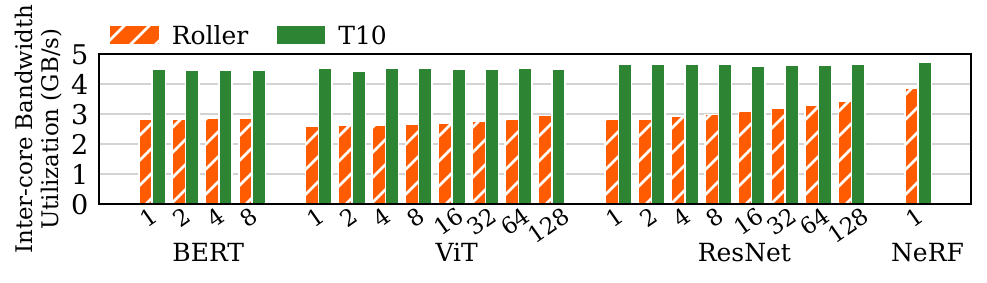}
    \vspace{-3.5ex}
    \caption{\hlE{Average inter-core bandwidth utilized by each core when executing various DNN models on IPU.}}
    \label{fig:noc_util}
    \vspace{-2ex}
\end{figure}

% \vspace{0.2em}
\noindent
\textbf{Inference latency breakdown.}
% Despite reducing on-chip data duplication, the temporal mapping in \pname{} incurs inter-core data transfer overhead. 
% \pname{}'s inter-operator scheduling aims to reduce operator setup overhead with minimal extra memory pressure ($\S$\ref{sec:design:inter_op}). 
We break down the compute time and inter-core data transfer time of Roller and \pname{} in \Cref{fig:compute_shift_breakdown}. 
Compared with Roller, \pname{} reduces the inter-core communication overhead from 50\%--74\% to only 8\%--43\%. 
% \hl{by efficiently utilize the on-chip interconnection.}
For most models, while the spatial-reduction nature of common operators like MatMul intrinsically incurs significant inter-core data sharing, \pname{} reduces the communication overhead by smartly shifting the shared tensor data.
For ResNet and NeRF, \pname{} automatically minimizes the inter-core movements of their large input activation tensors,
by efficiently sharing the smaller convolution kernels or model weights across the cores, enabling even lower communication overhead.
Similar patterns and benefits are also observed in the large language models examined in $\S$\ref{sec:eval_llm}, where \pname{} prefers to share the smaller input tensors while keeping the huge model weights stationary on each core.

\noindent
\hlE{\textbf{Inter-core bandwidth utilization.}
To further explain \pname{}'s low communication overhead, we show the average inter-core bandwidth utilized by each core during inter-core data transfers in \mbox{\Cref{fig:noc_util}}. While \pname{} uses an average bandwidth of 4.42GB/s to 4.73GB/s per core when running different DNN models (5.5GB/s is the advertised roofline), Roller only utilizes 2.61GB/s to 3.87GB/s, due to the inter-core communication inefficiencies discussed in \mbox{\S\ref{sec:limit_existing_approach}}.
Notably, models that shift more data each step (e.g., NeRF) have higher utilization.}

% For convolution-intensive models (e.g., ResNet), compute time dominates as the small convolution kernels can easily fit into each core's local memory and do not need to be shifted.
% As batch size increases, it takes more time to shift the input activation tensors.
% For matrix multiplication-intensive models (e.g., BERT, ViT, and NeRF), the spatial-reduction nature of MatMul intrinsically requires more data to be shifted among cores.
% Hence, data transfer time is high for small batch sizes, when the compute intensity is low.
% As batch size increases, compute time grows faster than data transfer time, since higher compute intensity in each core outweighs 
% the higher data transfer overhead (see more details in $\S$\ref{sec:eval_sens_vary_core_mem}).

\begin{figure}[t]
    \centering
    \vspace{0.5ex}
    \includegraphics[width=0.99\linewidth]{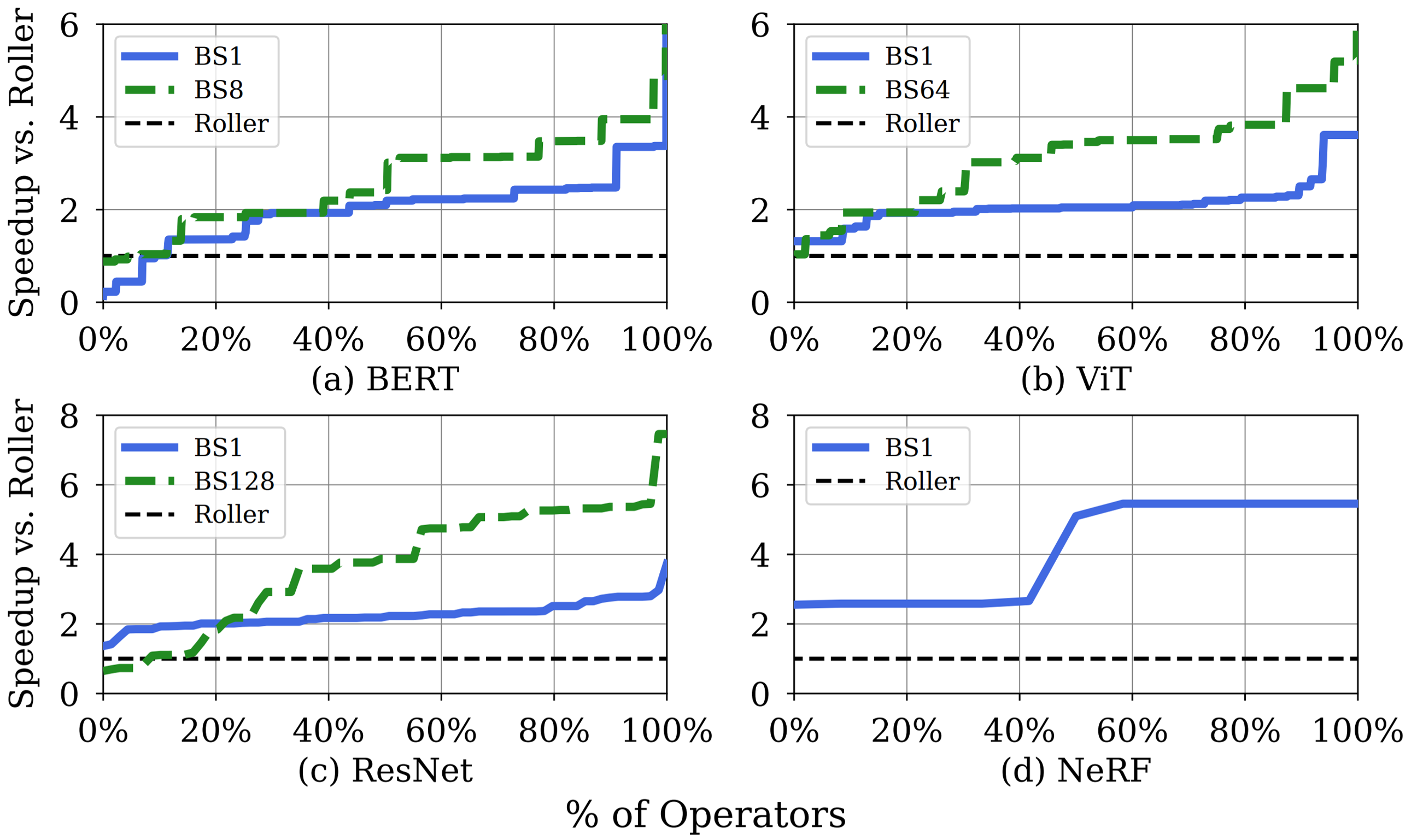}
    \vspace{-2.5ex}
    \caption{Distribution of \pname{}'s operator performance vs. Roller. We plot the min. and max. batch sizes as examples.}
    \label{fig:intra_op_time_cdf}
    \vspace{-1ex}
\end{figure}

% \vspace{0.2em}
\noindent
\textbf{\pname{} operator performance.} \pname{} improves the single operator performance by up to 10.79$\times$ (ResNet-BS8) compared to Roller. As \pname{} reduces the setup overhead of performance-critical operators, its inter-operator memory reconciliation introduces minimal negative impact on other operators (\S\ref{subsubsec:inter-operator}).
%As shown in \Cref{fig:intra_op_time_cdf}, 
\pname{} improves the performance of more than 80\% of the operators at the cost of slowing down less than 10\% (\Cref{fig:intra_op_time_cdf}).
% By searching for the best idle/active state memory size ratio, \pname{} 
% significantly outperforms Roller in end-to-end performance. 
%achieves significantly better end-to-end performance for DNN models than Roller.
% \Cref{fig:intra_op_time_cdf} shows \pname{}'s performance distribution of individual operators in each DNN model in comparison to Roller.

\begin{figure}[t]
    \centering
    % \vspace{0.5ex} % hotCRP compatible, do not change:
    \includegraphics[width=0.88\linewidth]{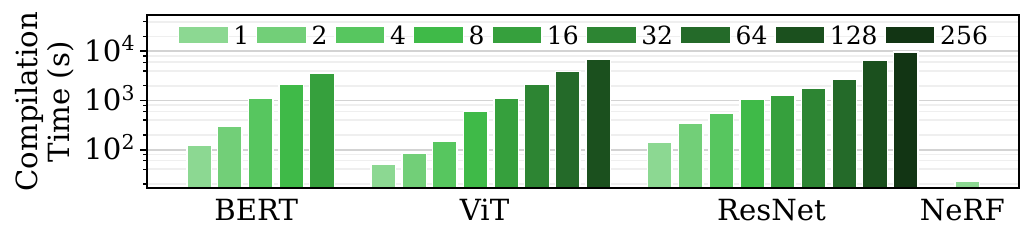}
    \vspace{-3ex}
    \caption{\pname{} compilation time of different batch sizes.}
    \label{fig:e2e_compilation}
    \vspace{-0.5ex}
\end{figure}

% \vspace{0.2em}
\noindent
\textbf{\pname{} compilation time.}
\Cref{fig:e2e_compilation} shows the compilation time of \pname{} for different models and batch sizes.
\pname{} finishes compilation in a few hours for most DNN inference programs (tested with AMD Ryzen 7950X3D).
As \pname{} exploits the predictability of the hardware architecture with the cost model and search constraints ($\S$\ref{subsubsec:intra-operator} and $\S$\ref{sec:implementation}), 
it avoids the expensive profiling for tuning each operator (e.g., Ansor~\cite{ansor}).

\begin{figure}[t]
    \centering
    % \vspace{-0.5ex}
    \includegraphics[width=0.9\linewidth]{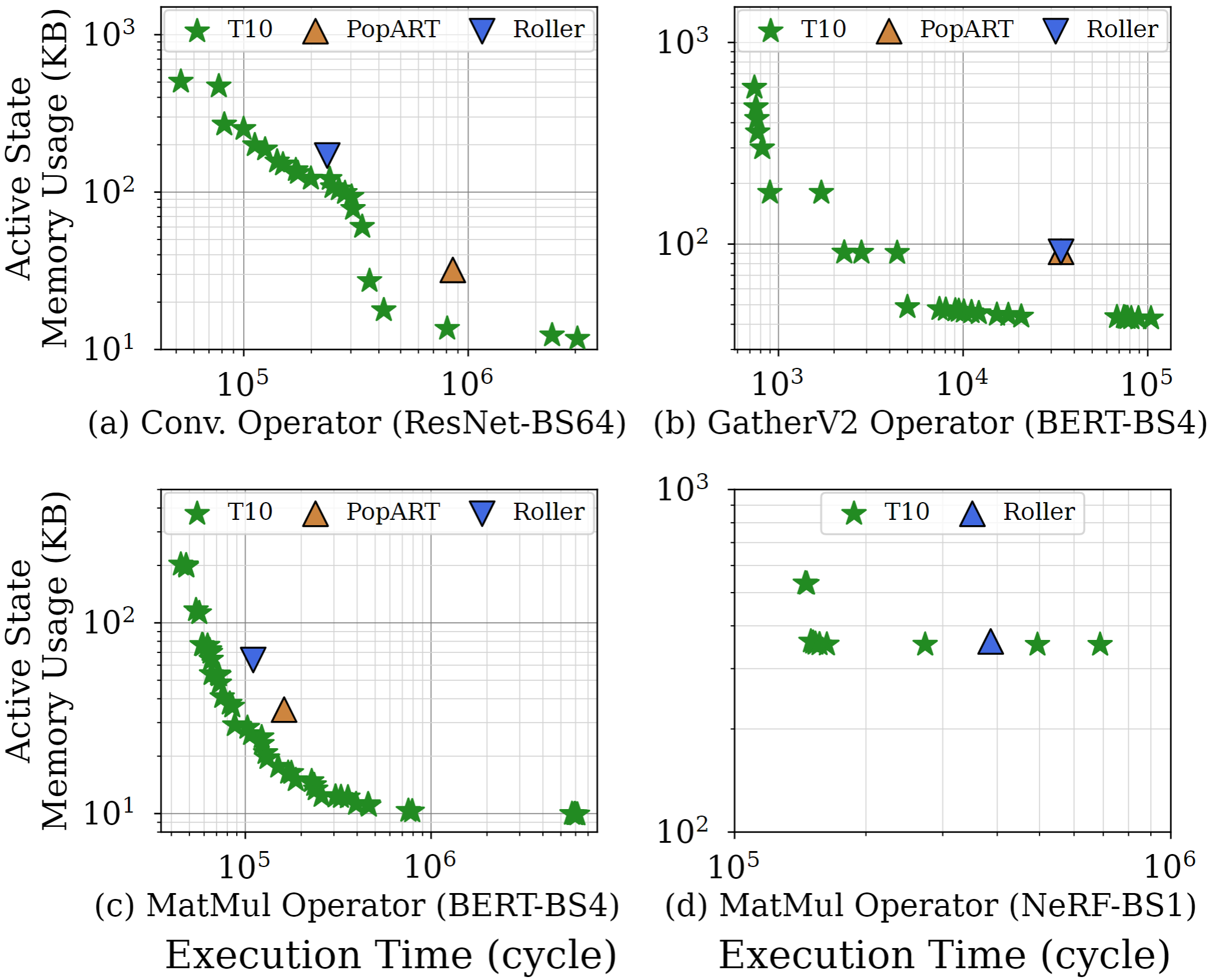}
    \vspace{-2ex}
    \caption{Candidate execution plans of representative operators (e.g., Figure (a) is a convolution operator from ResNet with batch size 32). Stars are optimal execution plans found by \pname{}. Triangles are the plans used by PopART and Roller, respectively. PopART fails to execute NeRF in Figure (d).}
    \label{fig:intra_performance}
    \vspace{-1ex}
\end{figure}

% \vspace{-3ex}
\subsection{Analysis of Intra-operator Optimization}\label{sec:eval_intra_op}
% \vspace{-0.3em}

We now examine \pname{}'s intra-operator optimizations ($\S$\ref{subsubsec:intra-operator}).

%We now analyze the effectiveness of \pname{}'s intra-operator optimization as discussed in $\S$\ref{subsubsec:intra-operator}. 
%We investigate the optimization process of individual operators and show how \pname{} finds the optimal candidate execution plans for each operator.

%\vspace{0.2em}
\noindent
\textbf{Flexibility of intra-operator plan selection.}
\pname{}'s intra-operator scheduling facilitates flexible inter-operator scheduling.
\Cref{fig:intra_performance} shows the set of optimal execution plans found by \pname{} for the representative operators compared to the plans used by PopART and Roller. The selected operators represent the majority of computation in a DNN workload. 
%(e.g., Convolution and MatMul), or have large search spaces and many candidate partitioning plans (e.g., GatherV2).

For most operators, \pname{}'s search space always contains an execution plan that is both faster and more memory efficient than PopART. Roller always tries to find the fastest plan that utilizes the most per-core local memory. However, Roller's maximum memory usage is limited due to the virtual global memory region (\Cref{fig:save_mem}). By removing the global memory, \pname{} enables a larger active memory and allows faster plans. Roller cannot make a globally optimized plan that exploits the trade-off between operators.
% \yuqi{TODO: With the same memory usage, \pname{} outperforms Roller by \hl{XX}$\times$ and \hl{XX}$\times$ for the representative operators due to less data duplication. To achieve the same performance as Roller, \pname{} requires \hl{XX}\% less memory capacity (\Cref{fig:intra_performance}d).}
% In contrast, by maintaining a set of Pareto-optimal plans, \pname{} enables the flexibility for the inter-operator scheduling stage to optimize for end-to-end performance.
In contrast, \pname{} maintains a set of Pareto-optimal plans and enables the inter-operator scheduling for optimized end-to-end performance.

% \hl{For example, \mbox{\Cref{fig:intra_performance}}c shows the sub-operator sizes (i.e., active state memory usage) under different partitioning plans for a MatMul operator. The top-left datapoint represents the partitioning plan with the least memory consumption (i.e., no duplication), but it suffers from long execution time. The bottom-right datapoint represents the partitioning plan with the largest memory consumption (i.e., duplicate as much as possible) and the fastest execution time. The difference in memory usage between them can be more than 20 times. Thus, it is important to emphasize that even if an operator is small enough to warrant duplicating shared weights on each core, partitioning over multiple cores still proves advantageous as it allows for memory savings to accommodate the weights from other operators on chip. Therefore, while PopART and Roller only consider one partitioning plan per operator, \pname{} considers all promising plans for each operator to identify different tradeoffs between sub-operator size and execution time.}

\begin{figure}[t]
    \centering
    \includegraphics[width=0.95\linewidth]{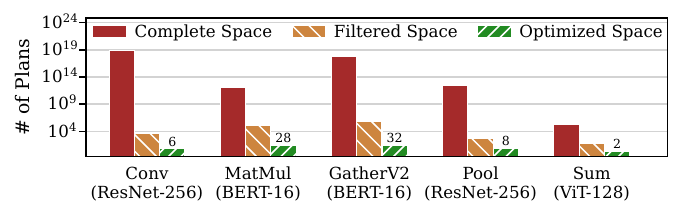}
    \vspace{-3.5ex}
    \caption{Intra-operator search space sizes. Each operator ``\texttt{Op (Model-BS)}'' is selected from \texttt{Model} with batch size \texttt{BS}.}
    \label{fig:intra_search}
    \vspace{-2.5ex}
\end{figure}

% \vspace{0.2em}
% \noindent
% \textbf{Cost model accuracy.} To test the accuracy of the linear regression cost models in \S\ref{sec:design:optimization}, we vary the shapes of the operator and compare the predicted execution time to the actual profiled execution time. \Cref{fig:intra_cost} shows the accuracy of some representative operator types. For most operators, \pname{} achieves near-perfect accuracy. The only exception is convolution, which is implemented with vendor-supplied kernels that apply some black-box optimizations. Even with slight inaccuracy, \pname{} can still find sufficiently good execution plans and outperform PopART and Roller. We envision that hardware vendors will supply a perfect cost model for their kernels as they integrate \pname{} in their toolchain in the future.

%  The configurations with more than 1472 cores are Vritual IPUs consisting of 2 or 4 IPU chips on the same board, connected by IPU-Link~\cite{v-ipu}. 

% \vspace{0.2em}
\noindent
\textbf{Intra-operator search space size reduction.}
To show how \pname{} explores the large search space of a single operator, we break down the search process and compare the remaining search space size after each critical step. \Cref{fig:intra_search} compares (1) the \textit{Complete Space} of all execution plans; (2) the \textit{Filtered Space} after applying the constraints defined in $\S$\ref{sec:implementation} but before applying the cost model; and (3) the \textit{Optimized Space} containing the Pareto-optimal plans selected by the cost model. 
\hlRA{In \mbox{\Cref{fig:intra_search}}, \texttt{Conv}, \texttt{MatMul}, \texttt{GatherV2} are the operators whose intra-operator optimizations generate the largest search space and contribute the most compile time, compared to other operators from the evaluated models in \mbox{\Cref{tab:benchmarks}}.}

\begin{figure}[t]
    \centering
    % \vspace{-1ex}
    \includegraphics[width=0.95\linewidth]{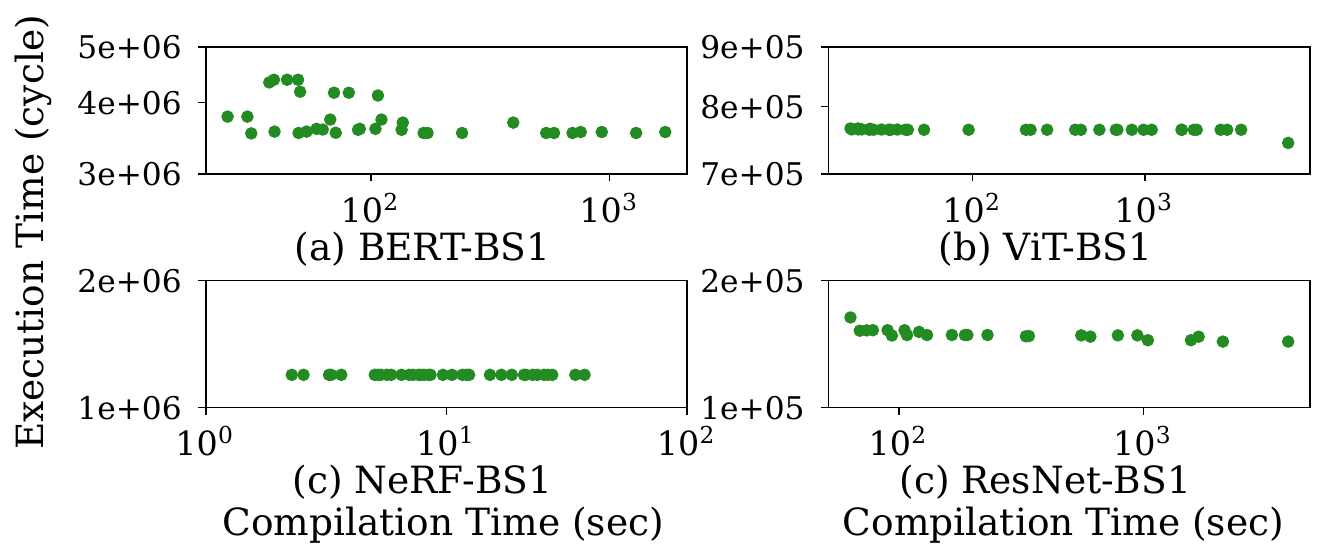}
    \vspace{-3ex}
    \caption{\pname{} compilation time and resulting execution latency with different constraint settings.}
    \label{fig:e2e_compilation_sens}
    \vspace{-2ex}
\end{figure}

\begin{figure}[t]
    \centering
    % \vspace{-0.5ex}
    \includegraphics[scale=0.22]{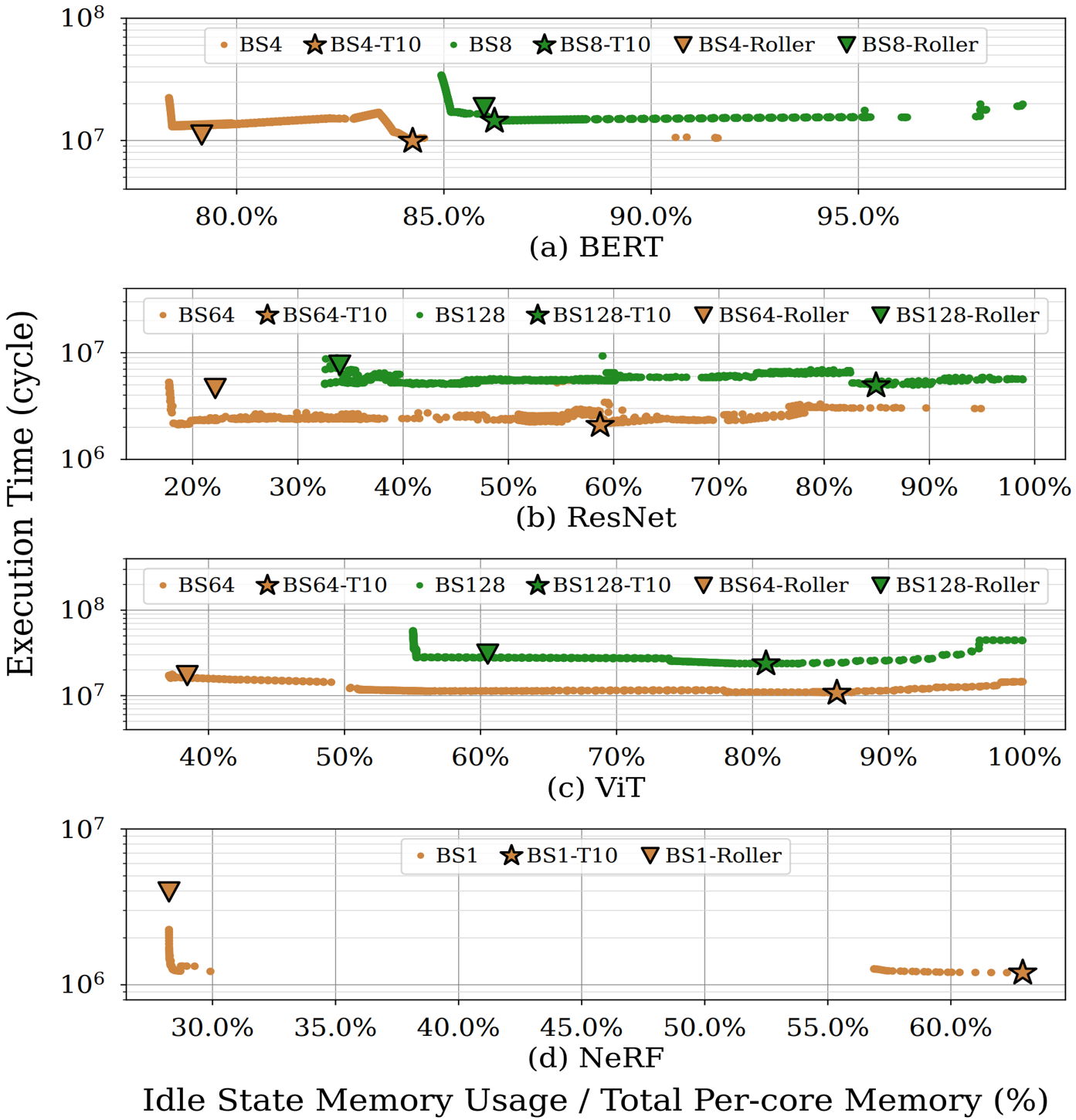}
    \vspace{-1.5ex}
    \caption{End-to-end execution plans. \texttt{BS\#} is batch size \texttt{\#}. Dots are plans explored by \pname{}. Stars and triangles are the final plans selected by \pname{} and Roller, respectively.} %\hl{TODO}}
    \label{fig:e2e_explore}
    \vspace{-1.5ex}
\end{figure}

The complete space grows exponentially with the number of dimensions in an operator
(\hlRA{up to $10^{19}$ plans for the largest convolution in ResNet,
which has 7 dimensions}).
The search constraints narrow down the search space to less than $10^4$ plans.
\hlC{As the efficient cost model in \mbox{\S\ref{subsubsec:intra-operator}} takes less than 100 milliseconds to evaluate each plan on one CPU core, \pname{} can explore $10^4$ plans in 30 seconds using 32 CPU cores.}
% and the plans in the filtered space are highly possible to be better than all other plans in the complete space. 
\hlRA{The final number of the Pareto-optimal plans is less than 50 for most operators,
and each operator's final plans can be cached and reused for identical operators within or across model(s).}
% The Pareto-optimal plans not only improve the single-operator performance but also provide enough flexibility for selecting the most appropriate plan during inter-operator scheduling.

% \vspace{0.2em}
\noindent
\textbf{Compilation time with different constraint settings.}
To compile faster, users can set stricter intra-operator search constraints (see $\S$\ref{sec:implementation}), which decreases the filtered space size and compilation time.
\Cref{fig:e2e_compilation_sens} shows the trade-off between execution performance and compilation time.
\hlRA{For most models, a strict constraint setting
that takes only one minute to compile, already yields near-optimal performance.}
% is sufficient, and relaxing the constraints has marginal benefit.

% As we relax the constraints, some DNN models (e.g., ResNet and ViT) can achieve slightly better performance.
% Note that BERT is sensitive to different constraint settings when the compilation time is small
% because different constraints have different impacts on the compilation time.

% \vspace{-3ex}
\subsection{Analysis of Inter-operator Optimization}\label{sec:eval_inter_op}
% \vspace{-0.3em}

We visualize the inter-operator search process by plotting the end-to-end execution plan explored at each search step in~\Cref{fig:e2e_explore}.
% \Cref{fig:e2e_explore} shows examples of the end-to-end execution plans explored and the final plans selected by \pname{}.
For most benchmarks, Roller generates the slowest (e.g., left-most) plan that requires the least idle state memory, which incurs significant operator setup time.
% This is because Roller only improves the active state memory usage for higher single-operator compute intensity without considering the idle-to-active state setup overhead.
This is because Roller does not optimize the idle-to-active state setup overhead of operators.
With inter-operator memory reconciliation ($\S$\ref{subsubsec:inter-operator}), \pname{} finds a globally optimized execution plan by recognizing the \textit{minimum active state memory demand}. For example, for ResNet-BS64 in \Cref{fig:e2e_explore} (b), \pname{} expands the idle state memory to about 58\% of the total on-chip memory by performing the setup phase for the performance-critical operators in advance. This improves end-to-end performance because many operators only demand at most 40\% of total memory as the active state memory.
% In contrast, neither Roller nor PopART is able to exploit this opportunity to better utilize on-chip memory.

% \begin{figure*}[t]
%     \centering
%     \includegraphics[width=0.76\linewidth]{figures/eval_sens_vary_core_mem.pdf}
%     \vspace{-3ex}
%     \caption{Performance of different IPU hardware configurations. \texttt{C(\#num\_cores)-M(\#size\_per\_core)} denotes \texttt{\#num\_cores} cores and \texttt{\#size\_per\_core} KB SRAM per core. NeRF is excluded since it cannot execute with smaller on-chip memory.}
%     \label{fig:vary_hardware}
%     \vspace{-2ex} 
% \end{figure*}

% \todo{compile time vs perf}

% \todo{intra-op heuristic search: minimize unused data amount, which is amount of operator data that is on-chip but not participating in computation. Example: two tensors are duplicated through one dimension, but only one is shifted (e.g., half of output in Figure5(c) does not participate in computation in each step.)}

% \todo{ipu does not always outperform GPU. Instead, it is more specialized for some domain. However, all accelerator designs with unique value worth the attention from the system community.}

\begin{figure*}[t]
    \centering
    % \vspace{-.5ex} % hotCRP compatible, do not change: -3.4ex scale=0.73
    \includegraphics[scale=0.206]{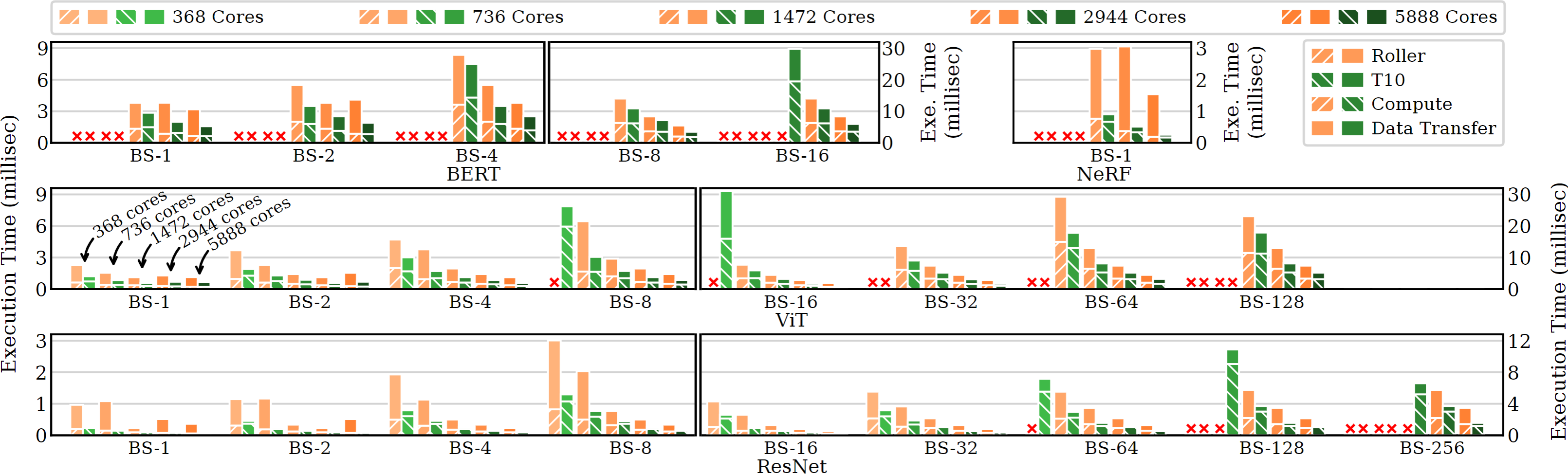}
    \vspace{-1.5ex}
    \caption{Performance of IPU with different number of cores. The \textcolor{red}{\ding{54}} indicates the DNN model cannot fit into the IPU chip.}
    \label{fig:scale}
    \vspace{-0.5ex} 
\end{figure*}

% \vspace{-2ex}
\subsection{Scalability of \pname{}}\label{sec:eval_sens_vary_core_mem}
% \vspace{-0.5em}

As the process node technology advances, we expect an intelligence processor will scale with more cores on a chip.
%To analyze the scalability of \pname{}, 
We emulate a larger chip by deploying \pname{} on a Virtual IPU (V-IPU)~\cite{v-ipu}, which exposes the 2 or 4 interconnected IPU chips on the same board to the compiler as a single IPU chip with 2,944 or 5,888 cores.
The inter-chip communication bandwidth is 160GB/s.
We also emulate smaller chips by restricting the number of cores in our compiler.
% We validate the results of larger chips against those of smaller chips.
% To validate emulation results, we also restrict the number of cores manually to evaluate \pname{} with fewer cores and compare the trend with that of our emulated chip.

\hlRA{\mbox{\Cref{fig:scale}} compares the execution performance of Roller and T10 on IPU devices with different numbers of cores.
\pname{} always outperforms Roller.}
% As shown in~\Cref{fig:scale},
% by using the inter-core connection and on-chip memory more efficiently, 
% \pname{} outperforms Roller regardless of the number of cores.
% The execution latency for these benchmarks with \pname{} on half-sized hardware is still more than 2$\times$ faster than with Roller on full-sized hardware. 
While both compilers achieve faster computation with more cores, 
\pname{} scales much better than Roller in terms of the inter-core data transfer, for two reasons.
First, the \rtensor{} abstraction eliminates the imbalanced and redundant inter-core data accesses caused by the VGM.
%Thus, the inter-core communication pattern of an \rtensor{} scales better than that of the virtual global memory. 
Second, \pname{}'s inter-core scheduling policy utilizes the on-chip memory more efficiently to decrease the data transfer volume and the setup time of large operators.
In most cases, \pname{} achieves similar or even better performance than Roller while using fewer cores. For example, \pname{} enables NeRF and ResNet with small batch sizes (BS-1 and BS-2) to be 2$\times$ faster than Roller while using only half of the cores.

With more than one chip, the inter-core communications in V-IPUs are bottlenecked by the inter-chip IPU-Link, causing the average effective inter-core bandwidth to drop by 26\%-33\%.
In this case, the execution time with Roller may even increase (e.g., ResNet BS-1) when we use more than one chip. In contrast, 
\pname{} does not increase the data transfer overhead despite the inter-chip communication overhead.
% because \pname{} can take advantage of the additional on-chip memory capacity to reduce the data transfer volume.

\begin{figure}[t]
    \centering
    \includegraphics[width=1\linewidth]{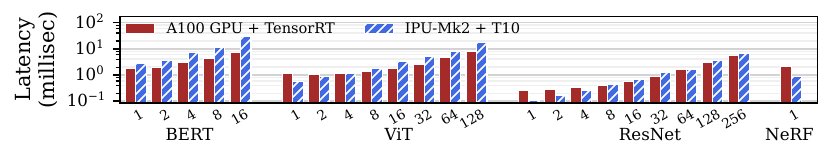}
    \vspace{-5ex}
    \caption{Inference latency of IPU+\pname{} vs. A100+TensorRT.}
    \label{fig:eval_gpu}
    \vspace{-1ex}
\end{figure}

% \vspace{-2.5ex}
\subsection{Comparison with A100 GPU}\label{sec:eval_gpu}
% \vspace{-0.5em}

To show how \pname{} unleashes the potential of inter-core connected architecture, 
% To study the benefits of the inter-core connected architecture, and show how \pname{} unleashes its potential, 
we compare an IPU MK2 chip against an A100 GPU, as both chips use 7nm technology and have similar FLOPS (\Cref{tab:compare_arch}). We run models with TensorRT~\cite{tensorrt} in an Azure NC24ads\_A100\_v4 instance.
We
% compile DNN models to
use FP16 and TensorCores on A100.
We warm up the models such that all data are in the HBM before running the experiments.

While \Cref{fig:eval_gpu} shows that TensorRT on A100 outperforms PopART, Ansor, and Roller on IPU in \Cref{fig:e2e_performance},
% However, 
\pname{} allows IPU to outperform A100 with small batch sizes by up to $2.44\times$.
% while the IPU chip only has 80\% of A100's peak TFLOPS (250 vs. 312 TFLOPS).
\hlF{\pname{} is especially good at small batches since it can efficiently utilize the fast on-chip memory of IPU, while the A100 is bottlenecked by loading data from off-chip memory.}
% \pname{} is especially good at small batch sizes, because its \rtensor{} abstraction and scheduling can efficiently utilize the fast on-chip memory of IPU, while many small operators on A100 are bottlenecked by loading data from off-chip memory. 

As batch size increases, the compute intensity increases, so the execution time becomes bounded by the peak FLOPS. Meanwhile, larger batch sizes also increase memory usage, leaving less space in the on-chip memory for \pname{} to trade off for less communication overhead. Due to the lower peak TFLOPS of IPU and the increased communication overhead when on-chip memory is nearly full, IPU suffers inferior performance in certain scenarios.
As intelligence processors continue to evolve, we believe they are a promising option for future latency-sensitive DNN workloads with \pname{} support.

\begin{figure}[t]
    \centering
    % \vspace{-1ex}
    \includegraphics[width=1.01\linewidth]{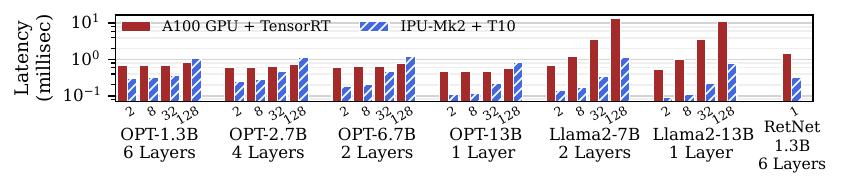}
    \vspace{-5ex}
    \caption{Inference latency of IPU+\pname{} vs. A100+TensorRT on large language models with different batch sizes.}
    \label{fig:opt_gpu}
    \vspace{-1ex}
\end{figure}

\begin{figure*}[t]
    \centering
    \vspace{-0.5ex}
    % \vspace{-3.3ex} % hotCRP compatible, do not change: -3.3ex width=0.99\linewidth
    \includegraphics[width=1\linewidth]{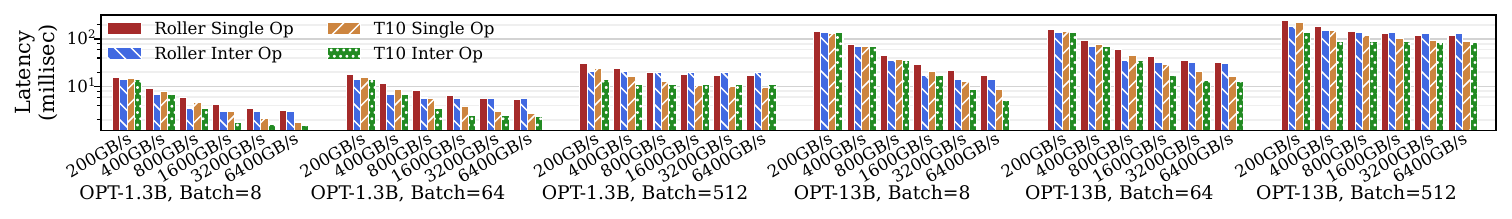}
    \vspace{-5ex}
    \caption{Emulated execution time of Roller and \oursys{} with different HBM bandwidths.}
    \label{fig:hbm}
    % \vspace{-0.5ex} 
\end{figure*}

% \vspace{-3ex}
\subsection{Performance of \pname{} for LLMs}
\label{sec:eval_llm}
% \vspace{-0.3em}

As Large Language Models (LLMs) become popular, inter-core connected DNN processors spot a new area to release their potential on high aggregated inter-core bandwidth. 
% In \mbox{\Cref{fig:opt_gpu}}, 
We examine two LLMs with standard transformer (OPT\mbox{~\cite{opt}} and Llama2\mbox{~\cite{llama2}}) and one LLM (RetNet\mbox{~\cite{retnet}}) with new transformer architecture designed for better scalability and memory-efficiency\mbox{~\cite{rwkv, retnet, mamba}}.
% Llama2~\cite{llama2}) and one retentive-network LLM (RetNet~\cite{retnet}) with different sizes on IPU using \pname{}. 
As the IPU chip we can access has insufficient on-chip memory to fit an entire LLM, we run a subset of layers for each LLM.
To run an entire LLM, users may follow common approaches that connect multiple chips as a pipeline\mbox{~\cite{bow_pod,groq-llm}}, where \pname{} reduces the latency and number of required chips by optimizing the execution and memory usage on each chip.
\hlF{The entire LLM's performance can be inferred from the single-chip performance:
since the intermediate data between layers is small (e.g., 131KB per token for Llama2-13B), the inter-chip communication overhead between pipeline stages is negligible.}
% This is because the intermediate data between layers is small , the inter-chip communication overhead between pipeline stages is negligible.}
% Thus, the entire LLM's performance is primarily decided by the single-chip performance.}

We compare the execution of LLM layers with A100 GPU in \mbox{\Cref{fig:opt_gpu}}.
Thanks to \pname{} and the fast inter-core connection, IPU achieves up to 16.38$\times$ lower inference latency (3.10$\times$ on average) than A100. 
By replacing virtual global memory with \rtensor{} abstraction (\S\ref{sec:limit_existing_approach}),
\pname{} frees up sufficient memory space for the large operators in LLMs.
A100 is limited by its 40MB global cache, which cannot fit a single large operator on chip. 
This requires A100 to load each model parameter from the slow off-chip memory multiple times.
Thus, the HBM bandwidth significantly bottlenecks the GPU performance for small batch sizes. 
For large batch sizes, similar to the case in \S\ref{sec:eval_gpu}, both GPU and IPU become compute-bounded, and IPU suffers from its lower peak FLOPS.

% \vspace{-1.5ex}
\subsection{Performance of \pname{} with Off-chip HBM}
\label{sec:eval_hbm}
% \vspace{-0.3em}

Given the absence of HBM on IPU, it is inefficient to serve LLMs on a single IPU chip, since it takes significant amount of time to load model parameters with off-chip memory (8GB/s). But if we combine the distributed on-chip memory with a large HBM, we can take advantage of both inter-core connection's high bandwidth and the HBM's large capacity.

We emulate HBM with different bandwidths on IPU by delaying the operator execution according to the predicted time of 
loading the operator from HBM using the roofline model~\cite{roofline}.
We extend \pname{} to support HBM.
We enable double buffering to overlap operator execution and HBM data transfer. The buffer size for execution/prefetching is 596MB/298MB, determined empirically based on the operator sizes. We examine two cases: (1) \texttt{Single Op}: we execute an operator and prefetch the next operator simultaneously; (2) \texttt{Inter Op}: we prefetch multiple operators as a group while the current group is executing. We apply \pname{}'s inter-op scheduling to decide the idle memory size for each operator.
%We decide which operators should be in the same group based on their tensor sizes, such that 
We ensure the minimum total memory requirement of the operator group is less than the prefetch buffer size. 
% When one operator is executing, the next operator is being loaded from HBM.
% Once the current operator completes execution and the next operator finishes loading, part of the memory space used by the completed operator is reserved to load the second next operator, a counter program starts to time the HBM load, and the already loaded operator starts execution.

For the \texttt{Single Op} case, Roller and \pname{} have similar performance when the HBM bandwidth is limited (Figure~\ref{fig:hbm}), as the execution is bottlenecked by the HBM.
When the HBM bandwidth increases, the execution becomes compute-bounded, \pname{} performs better, thanks to the execution plan identified by its intra-operator search.
% \pname{} performs better, thanks to the \rtensor{} abstraction and the efficient execution plan found by \pname{}'s intra-operator search, while Roller is bottlenecked by slower execution.
For the \texttt{Inter Op} case, both \pname{} and Roller perform better when the HBM bandwidth is low,
because grouping operators with different compute intensities helps balance the execution and prefetching.
With inter-operator scheduling, \pname{} achieves more benefits from grouping operators than Roller.
As we further increase the HBM bandwidth, the execution is compute-bounded, and \texttt{Inter Op} is slightly slower than \texttt{Single Op}, since operators in the same group compete for the on-chip memory capacity.

%% file: figures/table_compare.tex
% Please add the following required packages to your document preamble:
% \usepackage[table,xcdraw]{xcolor}
% If you use beamer only pass "xcolor=table" option, i.e. \documentclass[xcolor=table]{beamer}
\bgroup
\setlength\tabcolsep{2pt}
\begin{table}
\centering
% \caption{Comparison between different DNN accelerator architectures. The parameters shown are per chip.}
\caption{Hardware specifications (per-chip) of the A100 GPU and Graphcore IPU MK2 used in our evaluation.}
\vspace{-2.5ex}
% \todo{BSP differences not well shown}
\scriptsize
\begin{tabular}{|l|l|l|l|l|}

\hline                           & \textbf{A100 GPU~\cite{a100}}     & \textbf{IPU MK2~\cite{ipu2}} \\\hline 

% \textbf{On-chip Memory}           & Global-Local          & Local                 \\
% \textbf{Register Size (total)}            & 27MB      & 92KB       \\
\textbf{Local Cache (total)}        & 20.25MB   & 896MB    \\
% \textbf{Local Cache (per core)}  & 164KB & 624KB \\
\textbf{Global Cache}        & 40MB                  & N/A                 \\
% \textbf{Off-chip Memory}        & 80GB HBM                   & 16GB DDR4                \\
\textbf{Off-chip B/W}         & 2000GB/s              & 8GB/s                \\\hline                        
% \textbf{On-chip Transfer}         & In compute        & BSP                   \\\hline
\textbf{Inter-core B/W} & N/A & 6GB/s per link \\\hline
%\textbf{Off-chip Transfer}       & BSP/DMA         & BSP/DMA      & BSP                   \\\hline
\textbf{Number of Cores}           & 108                   & 1472                  \\
% \textbf{SIMD Width per Core}               & 2048bits              & 64bits  \\
\textbf{Total FP16 FLOPS}         & 312TFLOPS             & 250TFLOPS                    \\\hline
% \textbf{Core Config}             & \multicolumn{2}{c|}{SIMD Vector Unit + Systolic Array}        \\\hline

% \hline                           & \textbf{TPU v4~\cite{tpu_cloud}}       & \textbf{A100 GPU~\cite{a100}}     & \textbf{IPU mk2~\cite{ipu2}} \\\hline 

% \textbf{On-chip Memory}          & Global-Local          & Global-Local          & Local                 \\
% \textbf{Register Size (total)}           & Unknown               & 27MB      & 92KB       \\
% \textbf{Local Cache (total)}        & 32MB         & 20.25MB   & 896MB    \\
% \textbf{Global Cache}       & 256MB                 & 40MB                  & N/A                 \\
% \textbf{Off-chip Memory}       & 32GB HBM                   & 80GB HBM                   & 16GB DDR4                \\
% \textbf{Off-chip B/W}        & 1200GB/s              & 1555GB/s              & 32GB/s                \\\hline                        
% \textbf{On-chip Transfer}        & DMA        & In compute        & BSP                   \\\hline
% %\textbf{Off-chip Transfer}       & BSP/DMA         & BSP/DMA      & BSP                   \\\hline
% \textbf{Number of Cores}         & 2                     & 108                   & 1472                  \\
% \textbf{SIMD Width per Core}              & 4096bits              & 2048bits              & 64bits  \\
% \textbf{Total FP16 FLOPS}        & 260TFLOPS             & 312TFLOPS             & 250TFLOPS                    \\\hline
% \textbf{Core Config}             & \multicolumn{3}{c|}{SIMD Vector Unit + Systolic Array}        \\\hline

\end{tabular}

\label{tab:compare_arch}
\vspace{-1ex}
\end{table}
\egroup

%% file: discussion.tex
\section{Discussion and Future Work}\label{sec:discussion}

\noindent
\hlcommon{\textbf{Apply \pname{} to multiple chips.}
While \pname{} focuses on a single chip, it can be extended to optimize the inter-chip communication in modern AI infrastructures.
\pname{} can tradeoff between the inter-chip communication overhead and the per-chip memory consumption, enabling hardware clusters to execute larger DL workloads with higher performance.}
% \hlC{Also, the compute-shift paradigm of \pname{} can fit various hardware clusters. 
% For example, the rotation rings in \mbox{\S\ref{sec:design:compute}} can be easily mapped to TPU chips connected by a 3D-torus network\mbox{~\cite{tpu_v4_isca}}.}

\vspace{.5ex}
\noindent
\hlcommon{\textbf{Other inter-core connected hardware.}
The inter-core connection represents an important trend in AI chip evolution.
As large DL models are increasingly bounded by memory, new GPU architectures like Hopper\mbox{~\cite{h100}} also introduce inter-core links to connect the stream multiprocessors (SMs) into ``thread block clusters''.
More inter-SM data sharing allows less off-chip data access, which enables faster execution and consumes less energy.
% Also, as accessing on-chip memory costs less energy than accessing HBM, inter-core connections can improve energy efficiency with more on-chip data reuse.
\pname{} can be extended to optimize the inter-core communication in these new architectures.}

\vspace{.5ex}
\noindent
\hlD{\textbf{Combine IPU with HBM.}
Since the IPU device we can access is not equipped with HBM, we emulate HBM on IPU in \mbox{\S\ref{sec:eval_hbm}}. However, an inter-core connected chip recently released by SambaNova has HBM installed\mbox{~\cite{sn40l}}.
Similarly, IPU can support HBM by attaching HBM controllers to its on-chip interconnect, where HBM controllers can deliver data to cores in the same way as inter-core transfers.}

%% file: related.tex
\section{Related Work}
\label{sec:related}

\noindent
\hlcommon{\textbf{Deep learning compiler.}
DL compilers often focus on three design aspects: (1) intermediate representation (IR), which represents a computation workload; (2) computation model, which maps the computation to hardware; and (3) optimization, which identifies and explores the optimization space.}

\hlcommon{\pname{} uses the same IR with existing DL compilers: it uses operator graph \mbox{\cite{onnx}} to represent a DNN, and uses tensor expression \mbox{\cite{tensor_comprehensions}} to represent each operator.
As existing DL compilers \mbox{\cite{tvm, ansor, roller, xla, maestro, Triton, Mosaic}} are designed for the shared memory architecture, they use the ``load-compute-store'' computation model.
To explore this model's optimization space, they deploy techniques like tiling \mbox{\cite{tvm, maestro, ansor, roller, xla}},
loop reordering \mbox{\cite{maestro, tvm}},
polyhedral model \mbox{\cite{tvm, xla}},
load-compute overlapping \mbox{\cite{Triton}},
and sparsity \mbox{\cite{Mosaic}}.
% where tensor data is loaded from the memory hierarchy and computed, and the results are stored back to the shared memory.
As \pname{} targets a new distributed memory architecture, it uses a new ``compute-shift'' model and explores the new optimization space with new techniques.
\pname{} may work in orthogonal with other existing optimization techniques originally developed on the shared memory architecture, such as kernel fusion \mbox{\cite{xla, astitch, GraphTurbo:osdi2023}}, parallel kernel packing \mbox{\cite{rammer}}, and iteration batching \mbox{\cite{orca, friendliai}}.}

% Most DL compilers target conventional shared memory architectures~\cite{tvm, ansor, roller, xla, rammer}. 
% TVM~\cite{tvm}, Ansor~\cite{ansor}, and Roller~\cite{roller} focused 
% on intra-operator optimization by identifying the best tiling configuration for data reuse. 
% While Roller~\cite{roller} supports distributed on-chip memory, it still relies on the shared memory abstraction. 
% In contrast, \pname{} introduces a new abstraction for inter-core communications and provides an end-to-end compilation solution. 
% Other compilers like XLA~\cite{xla}, Rammer~\cite{rammer}, Astitch~\cite{astitch}, and GraphTurbo~\cite{GraphTurbo:osdi2023} optimized the interplay of 
%  individual operators, but they also target global shared memory architecture. 
% Their common techniques such as kernel fusion~\cite{xla, astitch, GraphTurbo:osdi2023} and parallel kernel packing~\cite{rammer} are complementary to \pname{}.
% \hlC{\pname{} is also compatible with other DL serving optimizations like iteration batching (e.g., Orca \mbox{\cite{orca}}), by generating one execution plan for each pre-configured batch size.}

\vspace{0.5ex}
% \vspace{1ex}
\noindent
\hlcommon{\textbf{Distributed model partitioning.}
% \vspace{0.5ex}
Many DL frameworks partition computation over distributed nodes with model parallelism, such as JAX \mbox{\cite{jax2018github}}, PartIR \mbox{\cite{partir}}, GSPMD \mbox{\cite{gspmd}}, Megatron \mbox{\cite{megatron}}, Alpa \mbox{\cite{alpa}}, FlexFlow \mbox{\cite{flexflow}}, and Tofu \mbox{\cite{tofu}}.
They adopt a single-program-multiple-data (SPMD) framework to shard an operator into multiple parallel sub-tasks, and insert a communication stage (e.g, AllReduce) when necessary to merge the sharded result.
The two stages are often optimized separately.
Also, to ensure each sub-task has complete local input data, some tensors will be replicated to multiple nodes, causing increased memory consumption.
In contrast, \pname{} schedules an operator as multiple interleaved compute and communication stages, and optimizes the stages holistically.}

% Many DL frameworks partition computation over distributed nodes~\cite{meshtf, matei:sosp2019, alpaserve:osdi2023, unity}.
% Tofu~\cite{tofu}, FlexFlow~\cite{flexflow}, Megatron~\cite{megatron}, and Alpa~\cite{alpa} 
% support general tensor parallelism by partitioning an operator or model along arbitrary tensor dimensions onto
% multiple GPUs or servers. 
% Unlike classical distributed systems that have large memory capacity but poor inter-node bandwidth, 
% our targeted architecture offers limited local memory on each core but high inter-core communication  
% bandwidth. \pname{} has different design trade-offs and goals. It aims to best utilize 
% the inter-core communications and distributed on-chip memory.  

% \vspace{1ex}
% \subsection{Dataflow and Systolic Architectures}
% \vspace{0.5ex}

\vspace{0.5ex}
\noindent
\hlcommon{\textbf{Dataflow architectures.}
Prior works facilitate the execution of various workloads on dataflow architectures.
% \cite{tenet, distal, plasticine:isca2017, amos:isca2022}. 
DISTAL \mbox{\cite{distal}} and Tenet \mbox{\cite{tenet}} provide scheduling primitives for users to write customized dataflow plans for linear algebra operations.
% Then, their compilers translate the user-provided execution plan into a distributed program.
SambaNova \mbox{\cite{plasticine:isca2017}} maps DNN execution to a dataflow accelerator using a mix of model and pipeline parallelisms, which may increase both latency (due to pipeline) and memory consumption (due to data replication).
Existing dataflow compilers \mbox{\cite{revet, spatial, systolic_general_1, systolic_general_2, systolic_general_3}} also focus on mapping general-purpose computation to interconnected CPU cores.
In contrast, \pname{} targets AI chips with thousands of fully connected tensor cores, and it can automatically compile an end-to-end DNN model into a compute-shift program with optimized execution latency.
The compute-shift model allows more generality and flexibility for diverse tensor operations than conventional hardware-defined systolic arrays \mbox{\cite{amos:isca2022, tpu_2017}}.}
% Also, compared to conventional systolic array-based studies \mbox{\cite{amos:isca2022, tpu_2017}}, \pname{}'s compute-shift model provides more generality and flexibility for diverse tensor operations.}

% Also, the compute-shift model provides more generality and flexibility than conventional systolic array-based studies \mbox{\cite{amos:isca2022, tpu_2017}}, by supporting diverse tensor operations on diverse hardware.}
% However, these works cannot be directly applied to optimize DNN models on IPU-like architecture.
% \hlC{Given massive multi-dimensional DNN operators and AI chips with thousands of fully connected tensor cores, \pname{} faces different opportunities, challenges, and tradeoffs.}

% Compared to conventional systolic array-based studies \cite{amos:isca2022, tpu_2017}, \pname{}'s compute-shift model provides more generality and flexibility for supporting diverse tensor operations on diverse hardware.
% First, it generalizes the systolic execution pattern to an N-dimensional torus beyond 2D systolic array.
% Second, it generalizes for arbitrary DNN workloads (e.g., tensor expressions with different logics, dimensions, and data shapes).
% Third, it enables a rich scheduling space (e.g., shift only 1/3 of the data in each step), so that compiler can automatically make the best optimization trade-offs.

%% file: conclusion.tex
\vspace{1ex}
\section{Conclusion}
\label{sec:conclusion}
% \vspace{-0.5em}

We present \pname{}, an end-to-end deep learning compiler for inter-core connected intelligence processors. 
%scaling inter-core communications with distributed on-chip memory on massive intelligence processors. 
%addressing the unique challenges posed by the absence of global shared memory and limited on-chip memory capacity. 
It generalizes the compute-shift computing paradigm on distributed on-chip memory for  
%\pname{} provides a systematic approach for 
enabling efficient operator partitioning and cost-aware operator scheduling. 
%in consideration of the trade-off between on-chip memory usage and inter-core communication overhead. 
We show its efficiency and scalability on a real massively-parallel intelligence processor. 
%distributed on-chip memory-based intelligence processor. 
%Our evaluation shows its superior 
%performance in comparison with state-of-the-art deep learning compiler techniques.

%By constructing a comprehensive spatial-temporal optimization space, 
%\pname{} enables efficient partitioning of operators into independent sub-operators mapped onto individual cores, 
%provides the flexibility to trade-off between memory usage and communication overhead, and employs a cost-aware operator 
%scheduling process to generate optimal execution plans. 

%and would sheld light on the development of scalable distributed memory-based accelerators. 
%The evaluation of \pname{} on various DNN models demonstrates its superior performance compared to state-of-the-art baselines, as well as its ability to handle much larger model and operator sizes. More importantly, \pname{} provides a systematic approach to potentially support larger-scale distributed memory accelerators in the future.

\vspace{1.5ex}

%% file: acknowledgment.tex
\begin{acks}
We thank the anonymous reviewers and our shepherd Byung-Gon Chun for their insightful comments and feedback. 
We thank the members in the Systems Platform Research Group at University of Illinois Urbana-Champaign for discussing and proofreading our paper. They include Benjamin Reidys, Haoyang Zhang, Shaobo Li, Daixuan Li, Eric Zhou, and Noelle Crawford. 
We thank Chen Jin, Chengshun Xia, and Han Zhao for the technical support on Graphcore IPU. 
This work was partially supported by NSF CAREER CNS-2144796.
\end{acks}